\documentclass{article}
\usepackage{cite}

\usepackage{arxiv}

\usepackage[utf8]{inputenc} 
\usepackage[T1]{fontenc}    
\usepackage{hyperref}       
\usepackage{url}            
\usepackage{booktabs}       
\usepackage{amsfonts}       
\usepackage{nicefrac}       
\usepackage{microtype}      
\usepackage{lipsum}
\usepackage[pdftex]{color, graphicx}
\usepackage{amssymb,amsmath}
\usepackage{color}

\begin{document}
	
\title{Wealth Distribution Models with Regulations: Dynamics and Equilibria}

\author{
	Ben-Hur Francisco Cardoso~\thanks{Instituto de Física, Universidade Federal do Rio Grande do Sul, 91501-970 Porto Alegre RS, Brazil.}\\
	ben-hur.cardoso@ufrgs.br\\
	\And
	Sebastián Gonçalves~\footnotemark[1]\\
	sgonc@if.ufrgs.br\\
	\And
	Jos\'e Roberto Iglesias~\footnotemark[1] ~\thanks{Instituto Nacional de Ci\^encia e Tecnologia  de Sistemas Complexos, CBPF, Rio de Janeiro, Brazil.}\\
	joseroberto.iglesias@gmail.com\\
}

\maketitle

\begin{abstract}
\noindent Simple agent based exchange models are a commonplace in the study of wealth distribution in an artificial economy. Generally, in a system that is composed of many agents characterized by their wealth and risk-aversion factor, two agents are selected sequentially and randomly to exchange wealth, allowing for its redistribution. Here we analyze how the effect of a social protection policy, which favors agents of lower wealth during the exchange, influences stability and some relevant economic indicators of the system. On the other hand, we study how periods of interruption of these policies produce, in the short and long term, changes in the system. In most cases, a steady state is reached, but with varying relaxation times. We conclude that regulations may improve economic mobility and reduce inequality. Moreover, our results indicate that the removal of social protection entails a high cost associated with the hysteresis of the distribution of wealth. Economic inequalities increase during a period without social protection, but also they remain high for an even longer time and, in some extreme cases, inequality may be irreversible, indicating that the withdrawal of social protection yields a high cost associated with the hysteresis of the distribution of wealth.
\end{abstract}

\section{Introduction}
Empirical studies of the distribution of wealth of people, companies and countries were first presented, more than a century ago, by Italian economist Vilfredo Pareto. He asserted that in different European countries and times the distribution of  wealth follows a power law behavior, i.e.  the cumulative probability $P(w)$ of agents whose wealth is at least $w$ is given by $P(w) \propto w^{-\alpha}$~\cite{pareto1897cours}. Today, that power law distribution is known as Pareto distribution, and the exponent $\alpha$ is named Pareto index. However, recent data indicates that, even though Pareto distribution provides a good fit to the distribution of high range of wealth, it does not agree with observed data over the middle and low range of wealth.  For instance, data from Sweden and France~\cite{richmond2006review}, India~\cite{sinha2006evidence}, USA~\cite{klass2007forbes}, and UK~\cite{druagulescu2001exponential} are fitted by a log-normal or Gibbs distribution with a maximum in the middle range plus a power law for high wealth range. The existence of these two regimes may be justified in a qualitative way by stating that in the low and middle class the process of accumulation of wealth is additive, causing a Gaussian-like distribution, while in the high class the wealth grows in a multiplicative way, generating the power law tail~\cite{nirei2007two}.

Different models of capital exchange among economic agents have been proposed trying to explain these empirical data. Most of these models (for a review, see~\cite{patriarca2010basic}) consider an ensemble of interacting economic agents where two of them are sequentially and randomly chosen to exchange a fixed or random amount of wealth. Aiming to obtain distributions with power law tails, several methods have been proposed to introduce a multiplicative factor in the exchanges. Keeping the constraint of wealth conservation, a well known proposition is that each agent saves a fraction -- constant or random -- of their resources~\cite{sinha2003stochastic, chatterjee2004pareto, chakraborti2000statistical, patriarca2004statistical, iglesias2003wealth, iglesias2004correlation, scafetta2002pareto}. With this proposition in mind, each agent $i$ is characterized by a wealth $w_i$ and a risk-aversion factor $\beta_i$. 

Let assume an exchange of wealth between agents $i$ and $j$. Supposing that $i$ wins an amount of wealth from $j$, we have that
$$w_i^* = w_i + dw \>;\> w_j^* = w_j - dw,$$
where $w_{i(j)}^*$ is the wealth of the agent $i(j)$ after the exchange. In this article we work with two rules that determine the quantity $dw$ transferred from the loser to the winner: the {\it fair} and the {\it loser} rule~\cite{hayes2002, caon2007unfair}. The first one states that $dw=\min[(1-\beta_{i})w_{i}(t);(1-\beta_{j})w_{j}(t)]$, while in the second we have $dw = (1-\beta_j)w_j$. The first rule is called {\it fair} because the amount of wealth exchanged is the same regardless of who wins, so the richer agent accepts to risk part of its wealth~\cite{hayes2002}. On the other hand, in the {\it loser} rule, the richer agent puts a higher amount at stake, thus, the exchange is only likely in situations where agents do not know the wealth of the others~\cite{sinha2003stochastic}.

Numerical~\cite{sinha2003stochastic, iglesias2004correlation, caon2007unfair} and analytical~\cite{moukarzel2007wealth} results of some
 variations of {\it fair} rule indicate that one possible result of that model is condensation, 
{\em i.e.} concentration of all available wealth in just one or a few agents. 
To overcome this situation, different rules of interaction have been applied, for example increasing the probability of favoring the poorer agent in a transaction~\cite{iglesias2004correlation, scafetta2002pareto}, or introducing a cut-off that avoids interactions between agents below and above this cut-off~\cite{das2005analytic}. Here we choose the former approach, using a rule suggested by Scafetta~\cite{caon2007unfair,scafetta2002pareto,laguna2005economic}, where, in the exchange between the agents $i$ and $j$, the probability of favoring the poorer partner is given by:
\begin{equation}
\label{eq:sca}
p=\frac{1}{2}+f\times\frac{|w_{i}(t)-w_{j}(t)|}{w_{i}(t)+w_{j}(t)},
\end{equation}
and $f$ is a factor that we call {\it social protection} factor, which goes from $0$ (equal probability for both agents) to $1/2$ 
(highest probability of favoring the poorer agent). 
In each interaction the poorer agent has probability $p$ of earn a quantity $dw$, 
whereas the richer one has probability $1-p$. 
It is evident that the higher the difference of wealth, the higher the influence of $f$ in the probability; 
thus, $f$ is a good indicator of the degree of application of social policies of income distribution.
In what follows we compare the results for the two rules in terms of the indicators of inequality 
and economic mobility. It is important to know weather the results are stable or not; for this reason, 
we define an equilibration criteria for the system. 
Besides, due to the fact that the probability asymmetry is intended to model a public regulation to protect the poore agents, it is reasonable to assume the $f$ is not constant in time. 
Thus, we analyze the effects that periods of interruption of these policies may have, 
in the short and long term.

All the simulations here presented have been performed for systems of $N = 1000$ agents and results have been averaged over $1000$ samples. Initial conditions are that both wealth ($w_i$) and risk-aversion factor ($\beta_i$), are uniformly distributed in the $[0, 1]$ interval. We use here the $MCS$ (acronym of {\it Monte Carlo Step}) as the unit of time, defined as the minimum number of steps needed for all agents to be able to be selected (i.e., $N / 2$), and the system evolves for a time $ T $, multiple of that value and long enough to verify equilibrium.

\section{Time evolution with a constant social protection factor}

\subsection{Inequality}

The Gini index is a measure of the inequality of a distribution and is often used to measure wealth inequality by economists and statistical organizations. It is defined as the ratio of the area enclosed by the Lorenz curve of the distribution (or cumulative distribution function) and the area under the uniform distribution~\cite{gini1921measurement}. In an operational way we define the Gini index as~\cite{sen1997economic}:
$$G(t) = \frac{1}{2}\frac{\sum_{i,j}|w_i(t) - w_j(t)|}{N \sum_i w_i}.$$
Gini index varies between $0$, which corresponds to perfect equality (i.e. everyone has the same wealth), and $1$, that corresponds to perfect inequality (i.e. one person has all the wealth, while everyone else has zero wealth). Here we will use the Gini index to measure the degree of inequality, but also, as a dispersion measure, to determine the stability of the wealth distribution. 

\begin{figure}[!htb]
	\centering
	\includegraphics[height=4.45cm]{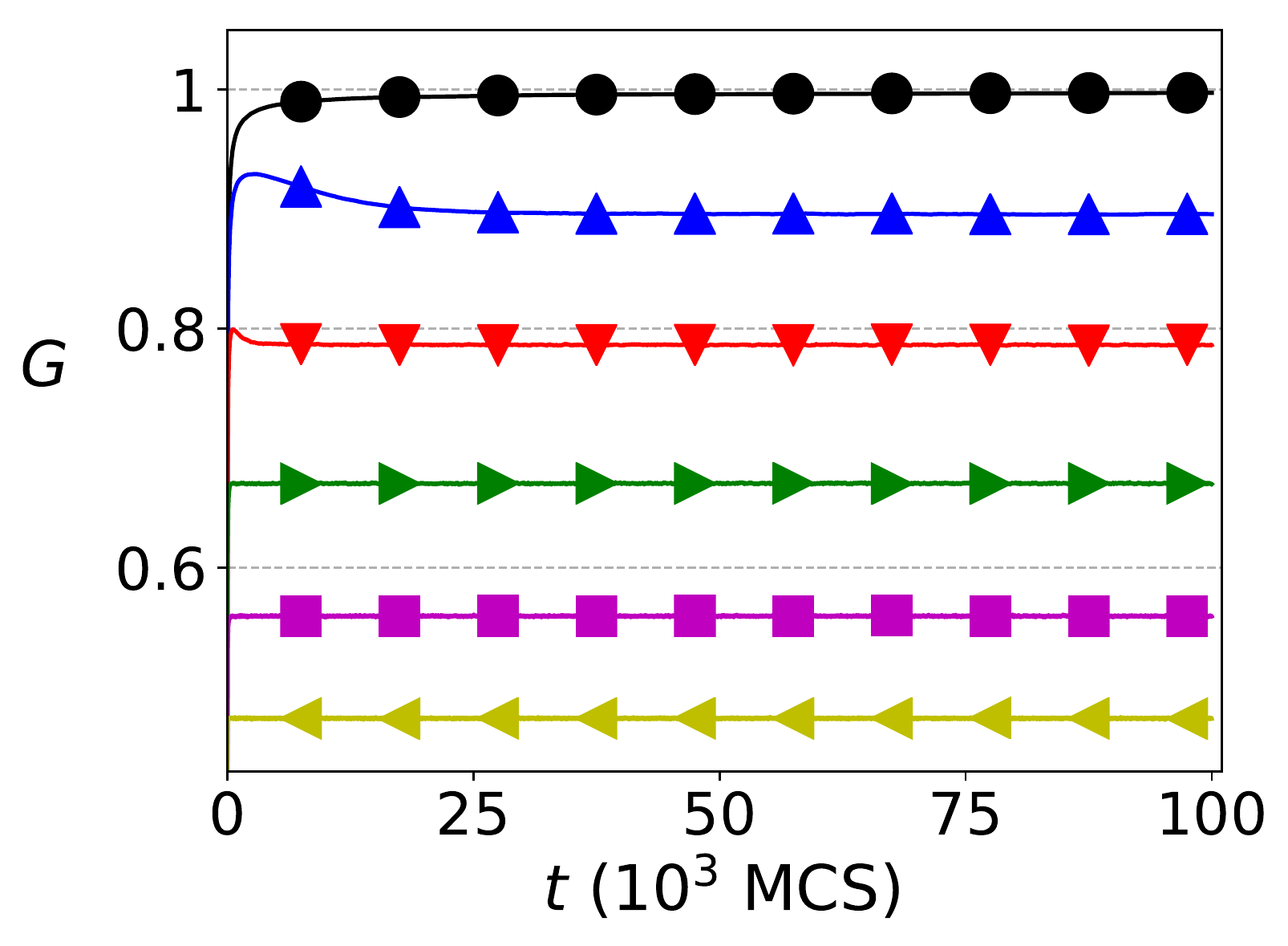}
	\includegraphics[height=4.45cm]{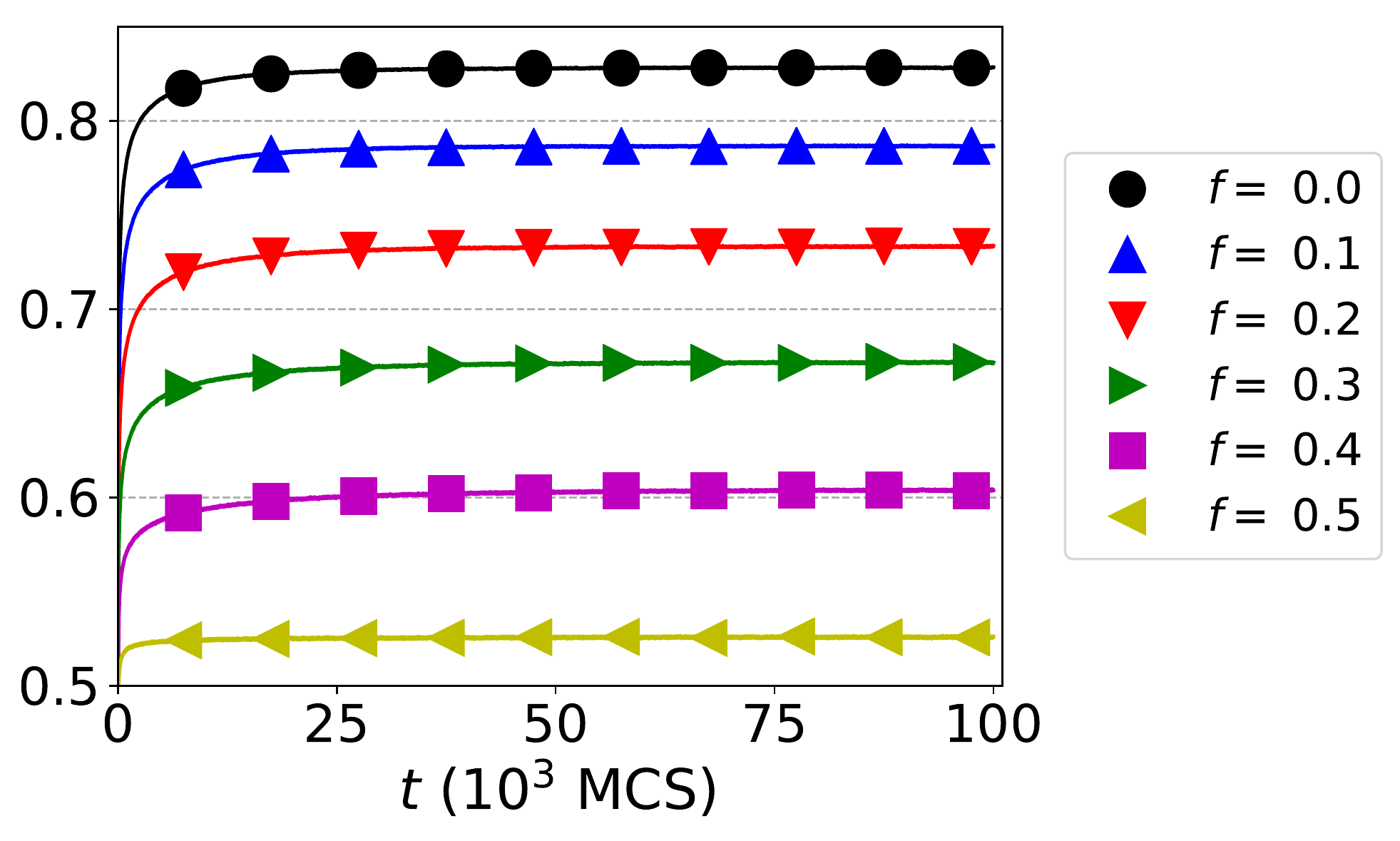}
	\caption{Time evolution of Gini index for different values of $f$, with {\it fair} (left) and {\it loser} (right) rule. The system arrives very fast to equilibrium for the {\it fair} rule, while it takes a  little longer for the {\it loser} rule. Notice that for $f=0$ the system arrives to condensation in the case of the {\it fair} rule, while it never condenses for the {\it loser} rule.}
	\label{fig:gini_time}
\end{figure}

\begin{figure}[!htb]
	\centering
	\includegraphics[height=4.45cm]{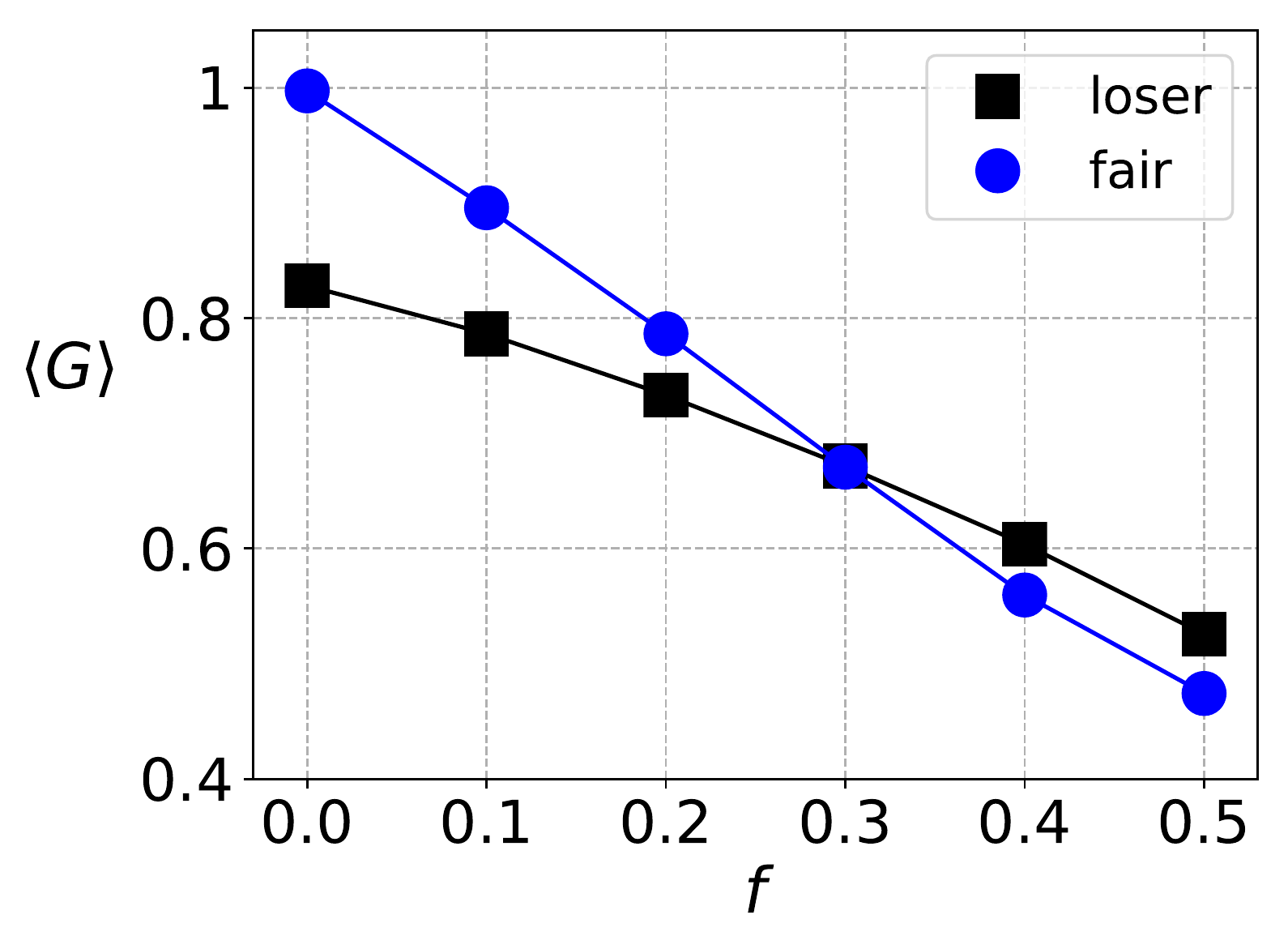}
	\caption{Average of Gini index as a function of $f$, for the two rules: {\it fair} and {\it loser}. Fair rule leads to higher inequality (and even condensation) for low values of the social protection factor, but to lower inequality when $f > 0.3$.}
	\label{fig:gini_comparison}
\end{figure}

Keeping this idea in mind, we show in Fig.~\ref{fig:gini_time} the time evolution of the Gini index, 
where we observe that generally it converges to a fixed value, after a transient of less than $50 \times 10^3 MCS$. 
For both rules, the higher the value of $f$, the smaller the inequality, as expected.
Also, we show in Fig.~\ref{fig:gini_comparison} the average Gini index over the last $10^4 MCS$ 
as a function of $f$, for both rules (error bars~\footnote{{Let be $A$ the last $n = 10^4 MCS$ of Gini index time series. The square of the error of $\mathbb{E}[A]$ is given by \cite{landau2014guide}: $\delta_A^2 = \frac{\sigma_A^2}{n}(1 +2\tau_A)$, where $\sigma_A^2 = \mathbb{E}[A^2] - \mathbb{E}[A]^2$ is the variance of $A$ and $\tau_A = \int_0^{\infty}\phi_A(t)dt$ is the autocorrelation time, where $\phi_A(t) = \frac{\mathbb{E}[A(0)A(t)] - \mathbb{E}[A]^2}{\sigma_A^2}$ is the stationary autocorrelation function.}} are smaller than the size of scatter plot). The {\it fair} rule leads to a greater inequality distribution in comparison to the {\it loser} rule, when $f < 0.3$.
For $f > 0.3 $, the opposite scenario is found; however the differences in Gini index on the right side of the crossing point are less significant than on the low $f$ region.

\begin{figure}[!htb]
	\centering
	\includegraphics[height=4.45cm]{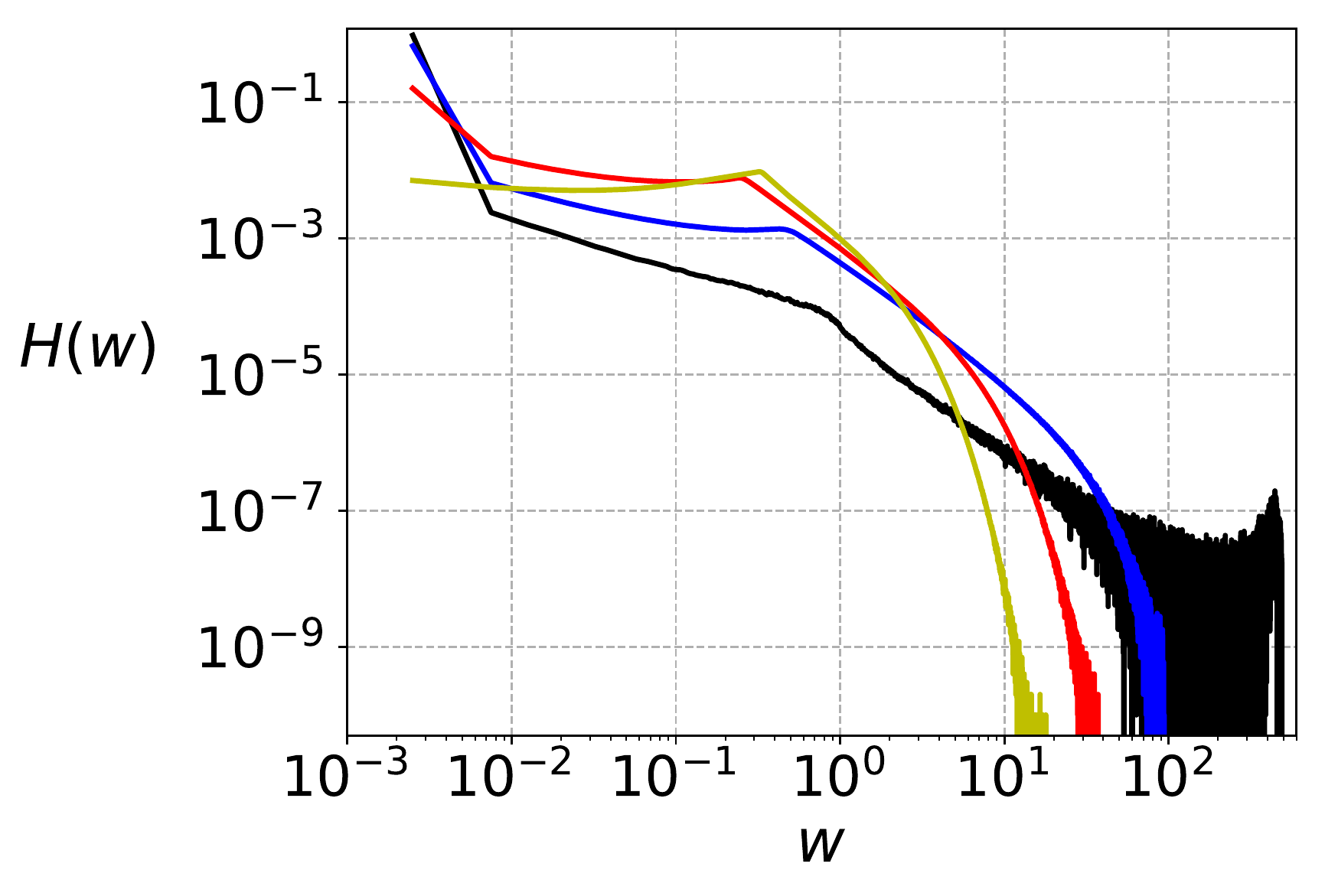}
	\includegraphics[height=4.45cm]{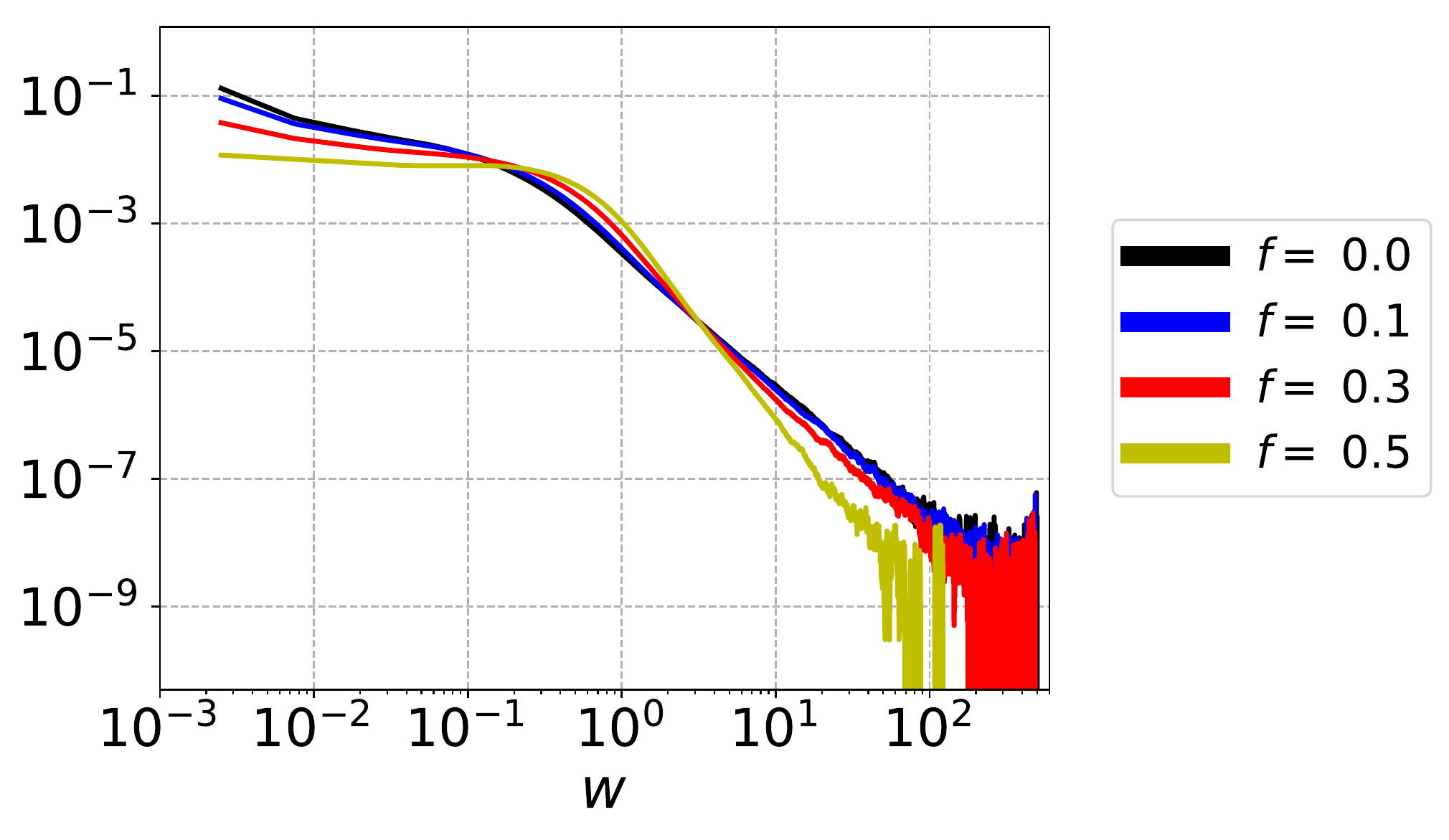}
	\caption{Equilibrium wealth distribution averaged over $10^4 MCS$ for different values of $f$, with {\it fair} (left) and {\it loser} (right) rule. Notice that for $f=0.3$ the distribution is different for the two rules, although having the same Gini index.}
	\label{fig:distribution}
\end{figure}

\begin{figure}[!htb]
	\centering
	\includegraphics[height=4.45cm]{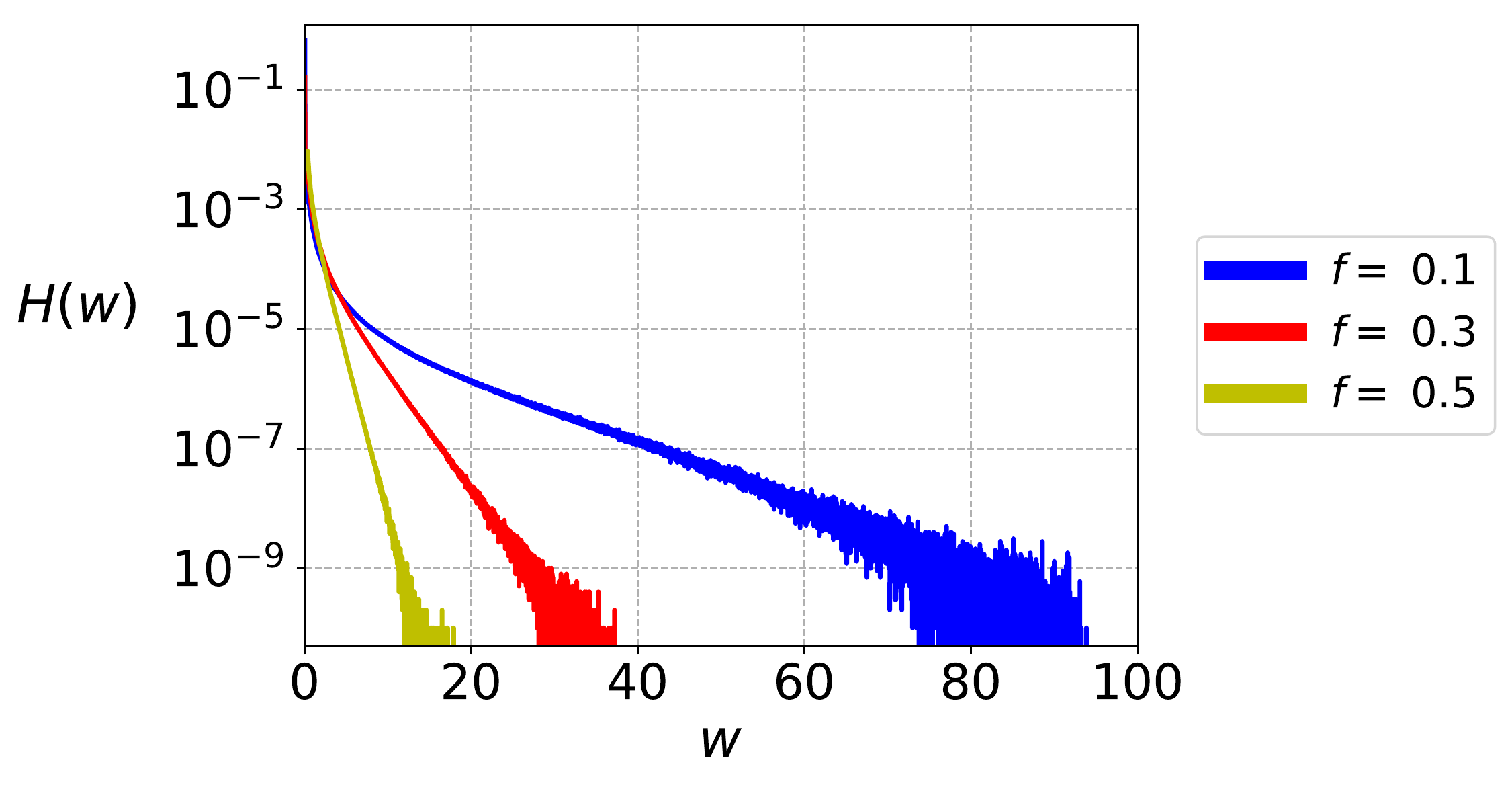}
	\caption{Equilibrium wealth distribution for the {\it fair} rule and for three different values of $f$. Distributions are the result of averages over $10^4 MCS$ and presented in semi-log scale. Notice that the distributions have a exponential tail.}
	\label{fig:distribution_zoom}
\end{figure}

As complementary information, we show in Fig.~\ref{fig:distribution} plots of the equilibrium wealth distribution, $H(w)$, for some selected values of $f$. $H(w)$ represents the fraction of agents that have wealth between $w$ and $w + dw$. We can see that there is an almost uniform region for the very low-wealth class, for both rules, if $f \lessapprox 0.3$. As for the high-wealth class, while the {\it loser} rule shows a power law tail with an $f$-dependent exponent (Fig.~\ref{fig:distribution}-right), the {\it fair} rule shows an exponential tail (Fig.~\ref{fig:distribution}-left and Fig.~\ref{fig:distribution_zoom}), except for $f = 0$, when a power law tail is observed.

Comparing the two rules, we notice that with $f < 0.3$, the Gini index from them diverge more (Fig.~\ref{fig:gini_comparison}), with the fair rule showing bigger inequalities. Such can be apparently contradictory with the distribution of Fig.~\ref{fig:distribution} where as we said there is an exponential tail in those cases, usually associated with lower inequalities than power law tail. However the bigger inequalities comes form the low-income region where for the fair rule an apparent tendency to condensation is observed as $f \to 0$.

\subsection{Economic Mobility}

Keeping the constraint of wealth conservation, economic mobility is only possible through exchanges. Due to this fact, a good indicator of mobility is the average wealth exchanged per unit time ($MCS$), called the {\it liquidity} of the system~\cite{iglesias2012entropy}, being defined as $$L(t) = \frac{1}{2N}\sum_i |w_i(t) - w_i(t-1)|.$$

\begin{figure}[!htb]
	\centering
	\includegraphics[height=4.45cm]{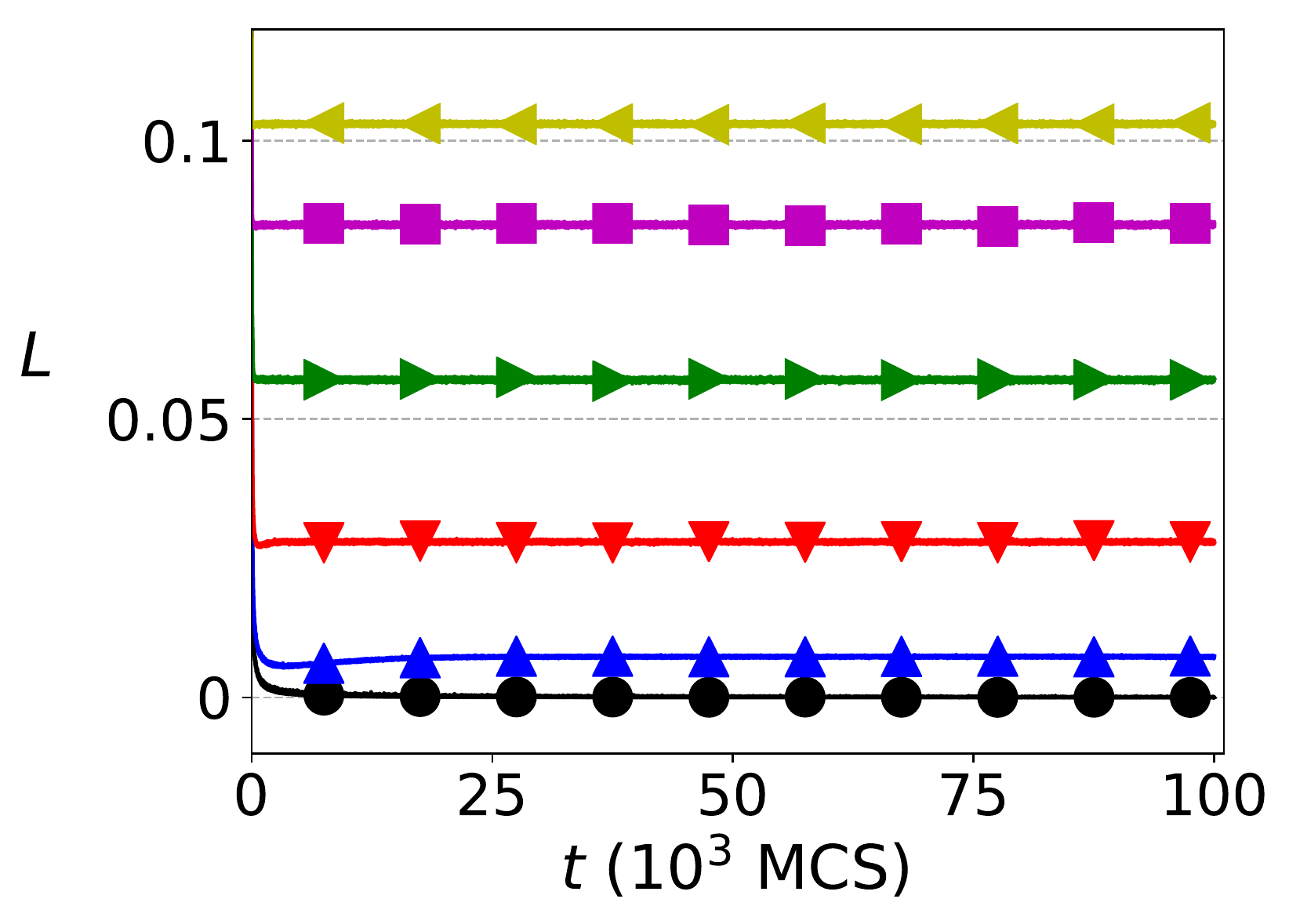}
	\includegraphics[height=4.45cm]{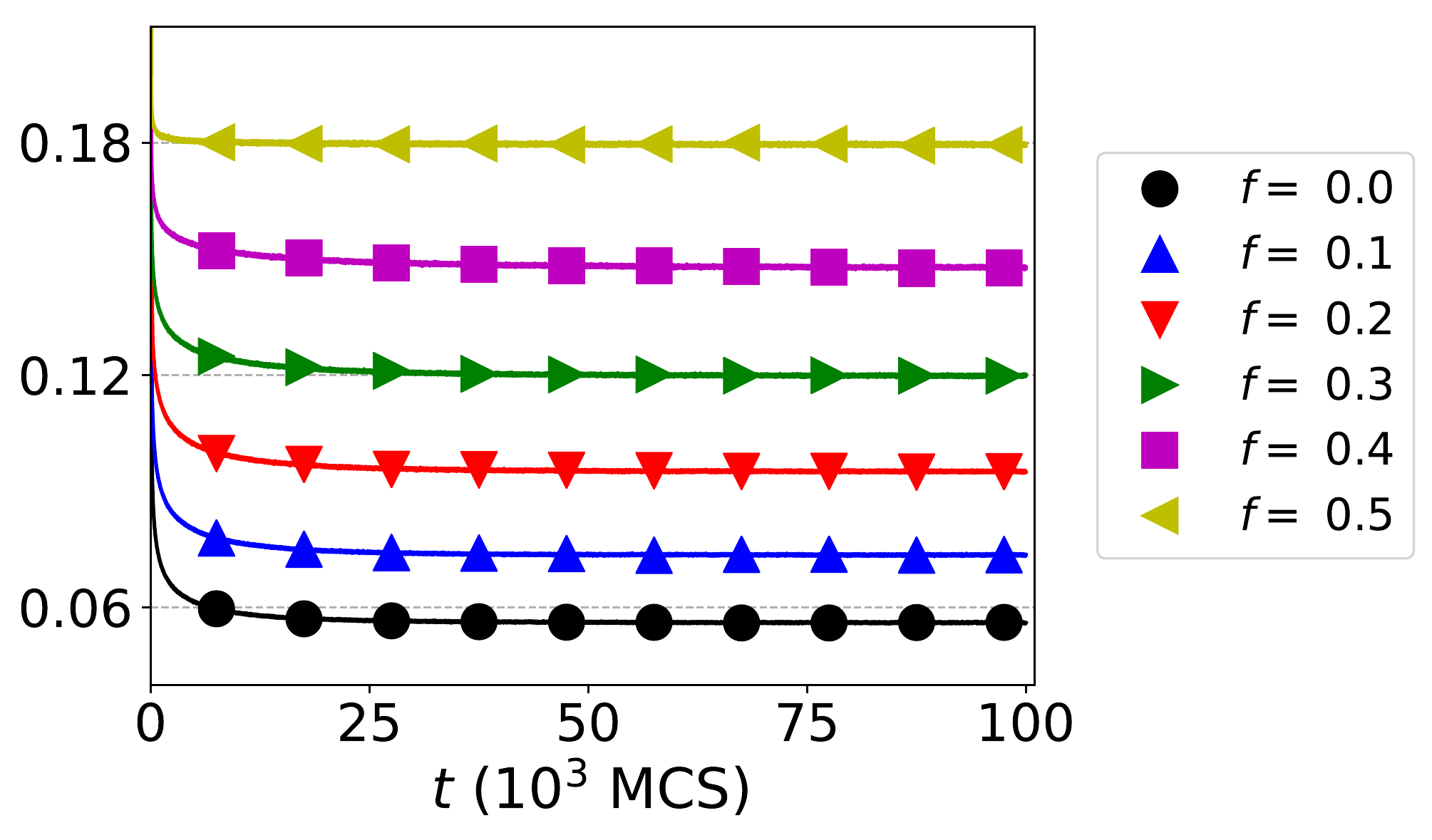}	
	\caption{Time evolution of liquidity for different values of $f$, with {\it fair} (left) and {\it loser} (right) rule. Liquidity may go to zero in the case of the {it fair} rule.}
	\label{fig:liquidity_time}
\end{figure}

\begin{figure}[!htb]
	\centering
	\includegraphics[height=4.45cm]{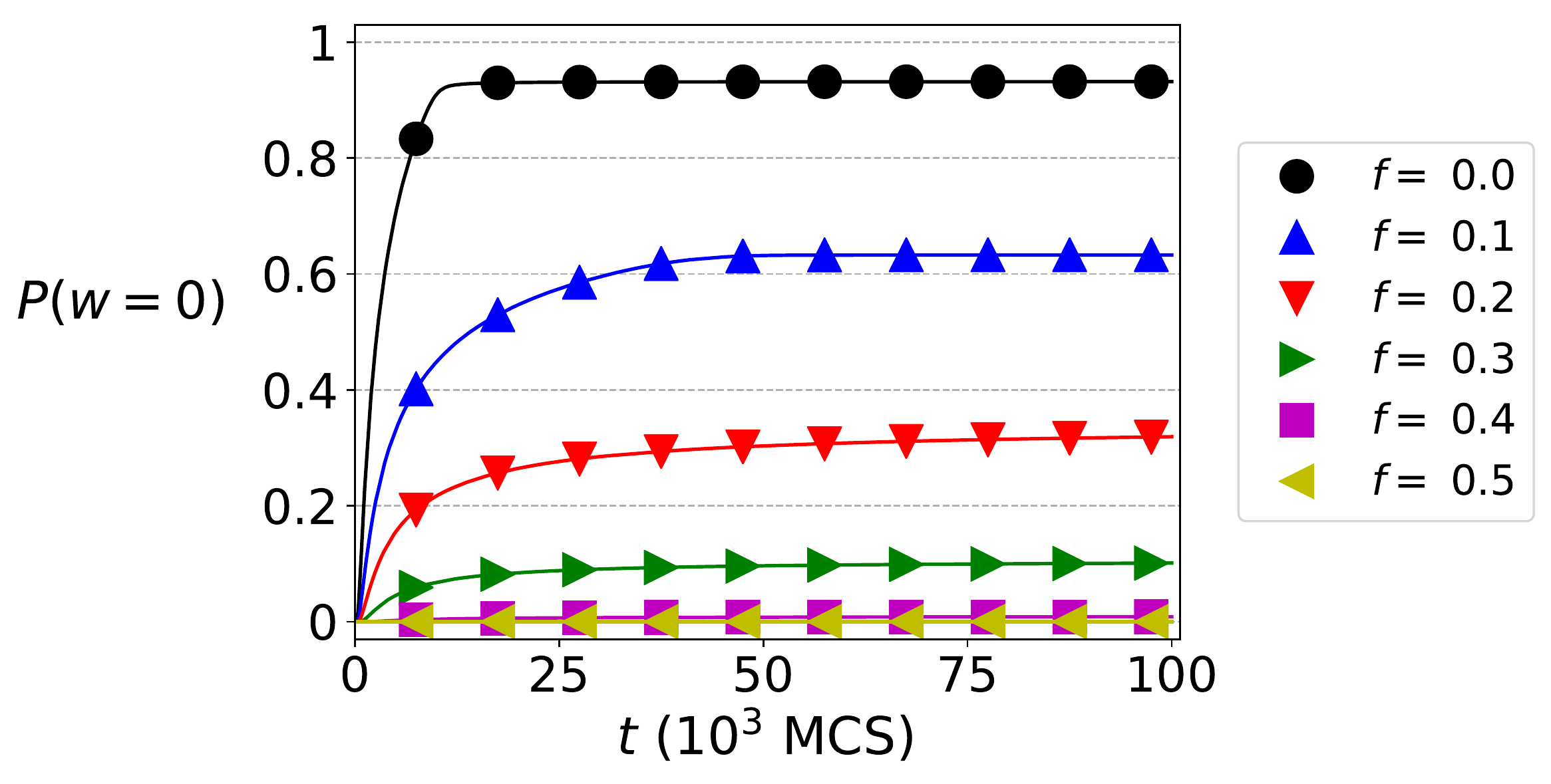}
	\caption{Fraction of agents with zero wealth for different values of $f$, with {\it fair} rule.}
	\label{fig:zero_time}
\end{figure}

We show in Fig.~\ref{fig:liquidity_time} the time evolution of the liquidity, where we can observe that, for both rules, the mobility increases with the increasing of value of $f$. Comparatively, the system with {\it fair} rule has less economic mobility. This is due to the restriction of this rule itself: only the minimum of the two risked wealth is exchanged, unlike the other rule, where the amount exchanged can be the full stake. Furthermore, as we show in Fig.~\ref{fig:zero_time}, in the case of the {\it fair} rule a significant fraction of agents (that decreases with increasing $f$) have zero wealth~\footnote{The global wealth in 2014 was less than $~10^{15}$ USD~\cite{lange2018changing}, while in our artificial system the total wealth is $500$. Since USD unit is discretized by the minimum value of $0.01$ USD, by analogy we set the minimum unit as $10^{-15}$. Therefore, we consider the wealth of an agent $i$ equal to zero if $w_i < 10^{-15}$.}; thus, whenever they are selected no wealth is exchanged, causing a lower mobility. It is interesting to observe that in the case of the loser rule the liquidity is never zero, keeping the economy at work. 

\subsection{Path to equilibrium}

Similar results are found in a previous work~\cite{caon2007unfair}, where the same system evolves for $10^4 MCS$. Moreover, it was shown in that work that for some low values of $f$ the stabilization was not complete, and a slow dynamics was still present. This can be explained now due to the lowest liquidity, showed in Fig.~\ref{fig:liquidity_time}. Here, systems evolve for periods ten times longer; therefore, it is worth to verify the stabilization of the system after this longer time. To do so, we use the time series of the Gini index: the system will be said to be equilibrated when a measure that depend on two times ( $t$ and $t + \tau$) is invariant under a temporal translation, that is, when it only depends on the interval $\tau$, and not on the time $t$ itself.

With this picture in mind, the system will be said to be equilibrated when the time series $G_t = \{G(t) \cdots , G(T)\}$ is second order stationary, that is, when $R(t, \tau) = R(\tau)$ for all $t$ , being $R(t,\tau) = \mathbb{E}[(G_t - \mathbb{E}[G_t])(G_{t+\tau} - \mathbb{E}[G_{t+\tau}])] $ the autocovariance function~\cite{chatfield2016analysis}. In an operational way, we divide the Gini index time series into $(T- M)$ time series of length $M$: $\{G(t) \cdots , G(t+M) \}$; so, the autocovariance function is estimated as
$$R(t,\tau) \approx \frac{1}{M} \sum_{n=1}^M[G(n+t) - \langle G \rangle _t][G(n+t+\tau)-\langle G\rangle_{t+\tau}]\>, \>\>\> \langle G \rangle _t = \frac{1}{M}\sum_{n=1}^MG(n+t).$$

This equilibrium condition implies that, for all $t$, $\frac{\partial R}{\partial t } = 0$. The distance from equilibrium for a given time series $G_t$ can be given by the mean square error ($MSE$) of this derivative over all associated values of $\tau$, that is, 
$$MSE(t) = \left \langle \left ( \frac{\partial R}{\partial t} \right )^2 \right \rangle_{\tau} \approx \frac{1}{M}\sum_{\tau=0}^M \left(\frac{R(t+1,\tau) - R(t-1,\tau)}{2} \right)^2.$$

\begin{figure}[!htb]
	\centering
	\includegraphics[height=4.45cm]{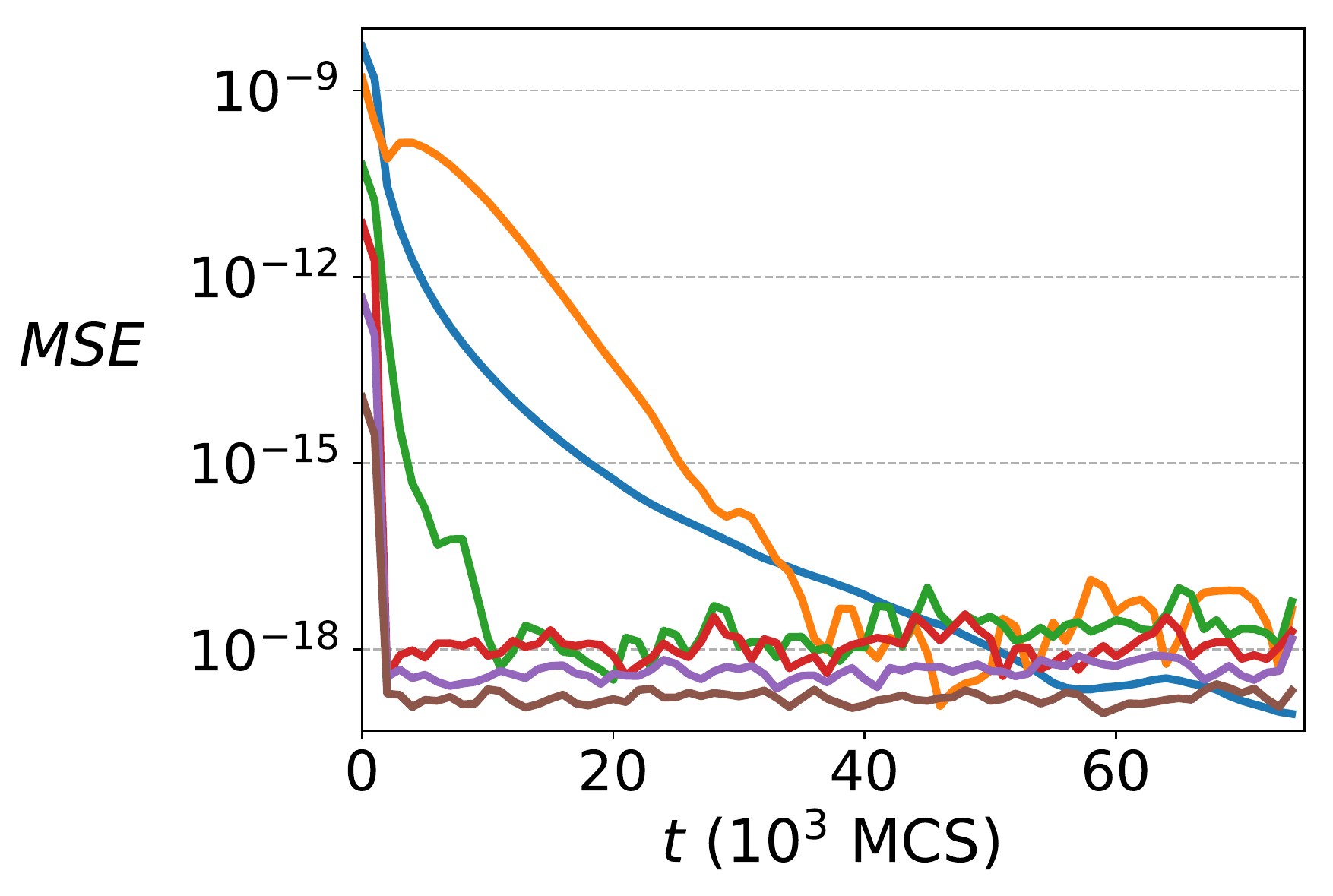}
	\includegraphics[height=4.45cm]{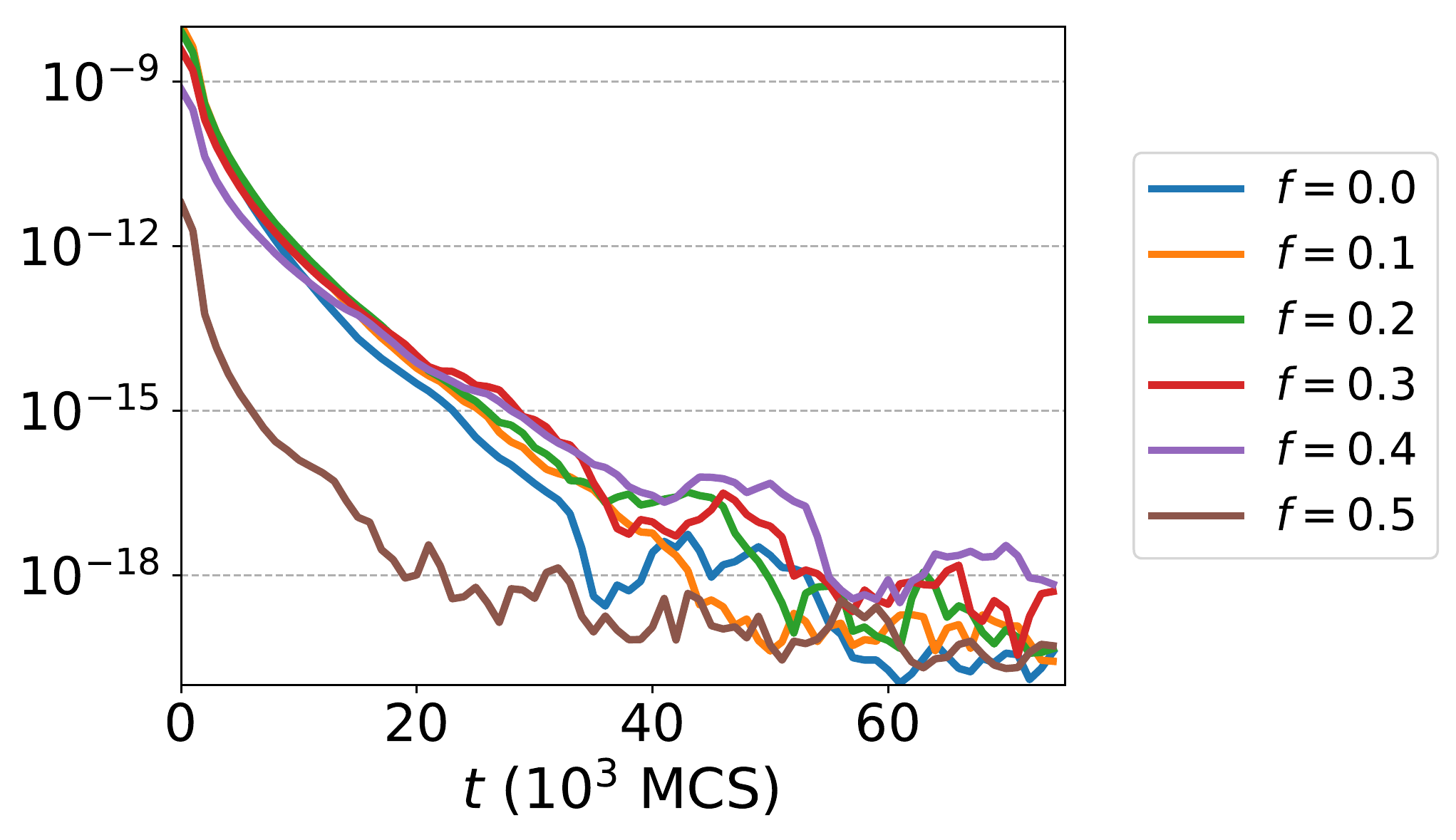}
	\caption{Mean square error ($MSE$) for different values of $f$, with {\it fair} (left) and {\it loser} (right) rule. $MSE$ is very small ($10^-9$) even for short times and arrives to extremely low values for $t$ of the order of $40 \times 10^3$ MCS.}
	\label{fig:mse}
\end{figure}

Working with $M = 25 \times 10^3 MCS$, we show in Fig.~\ref{fig:mse} the time evolution of this mean square error, where we can observe that, for both rules, the system arrives to an stable state.  

\newpage
\section{Perturbation of the social protection factor}
\begin{figure}[!htb]
	\centering
	
	\includegraphics[width=10cm]{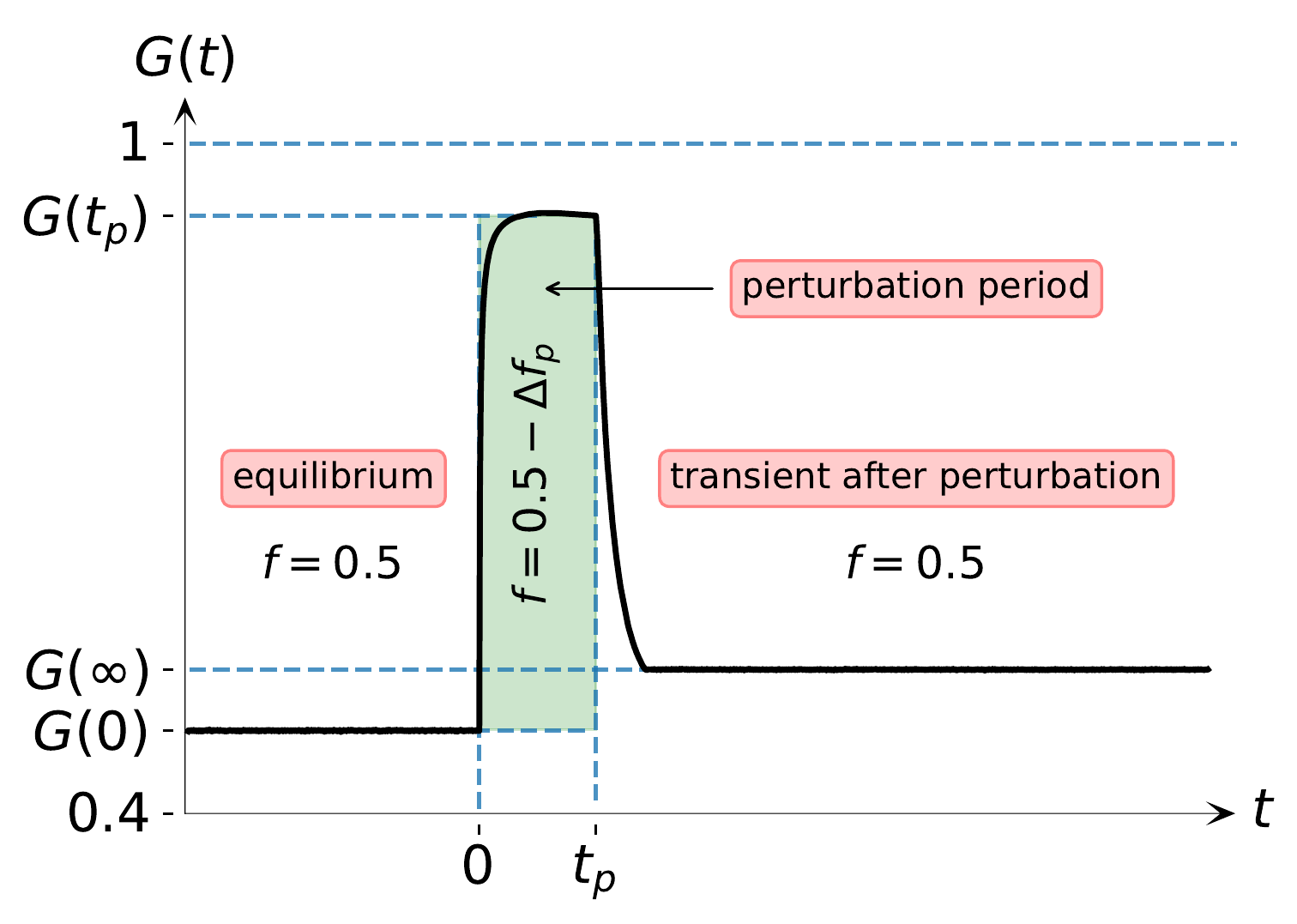}
	
	\caption{Schematic description of the perturbation procedure, exemplified  for  the {\it fair} rule, with $t_p = 6 \times 10^3 MCS$ and $\Delta f_p = 0.4$.}
	\label{fig:pert}
\end{figure}

In the previous section, we studied the effect of a constant social protection policy, {\em i.e.} a constant value of $f$. This factor summarize the set of regulation measures to protect the low income sector of the population. However, this kind of policies are subjected to political changes and so a constant value of $f$ is rather artificial. In this section, we analyze the consequences ---in the short and the long term--- of changes in social protection during a {\it perturbation period}. Particularly, we will focus on a common pratice nowadays, a reduction or elimination of social protection. To do so, we use the following approach: with the system initially equilibrated at $f = 0.5$, the value of $f$ is changed suddenly to $f = 0.5 - \Delta f_p$ and kept in this value during $t_{p}$ steps, where $\Delta f_p$ is called the {\it intensity} of the perturbation. After this perturbation period, the value of $f$ returns to its initial value $f = 0.5$, as schematically shown in Fig.~\ref{fig:pert}. 

\begin{figure}[!htb]
	\centering	
	\includegraphics[height=4.45cm]{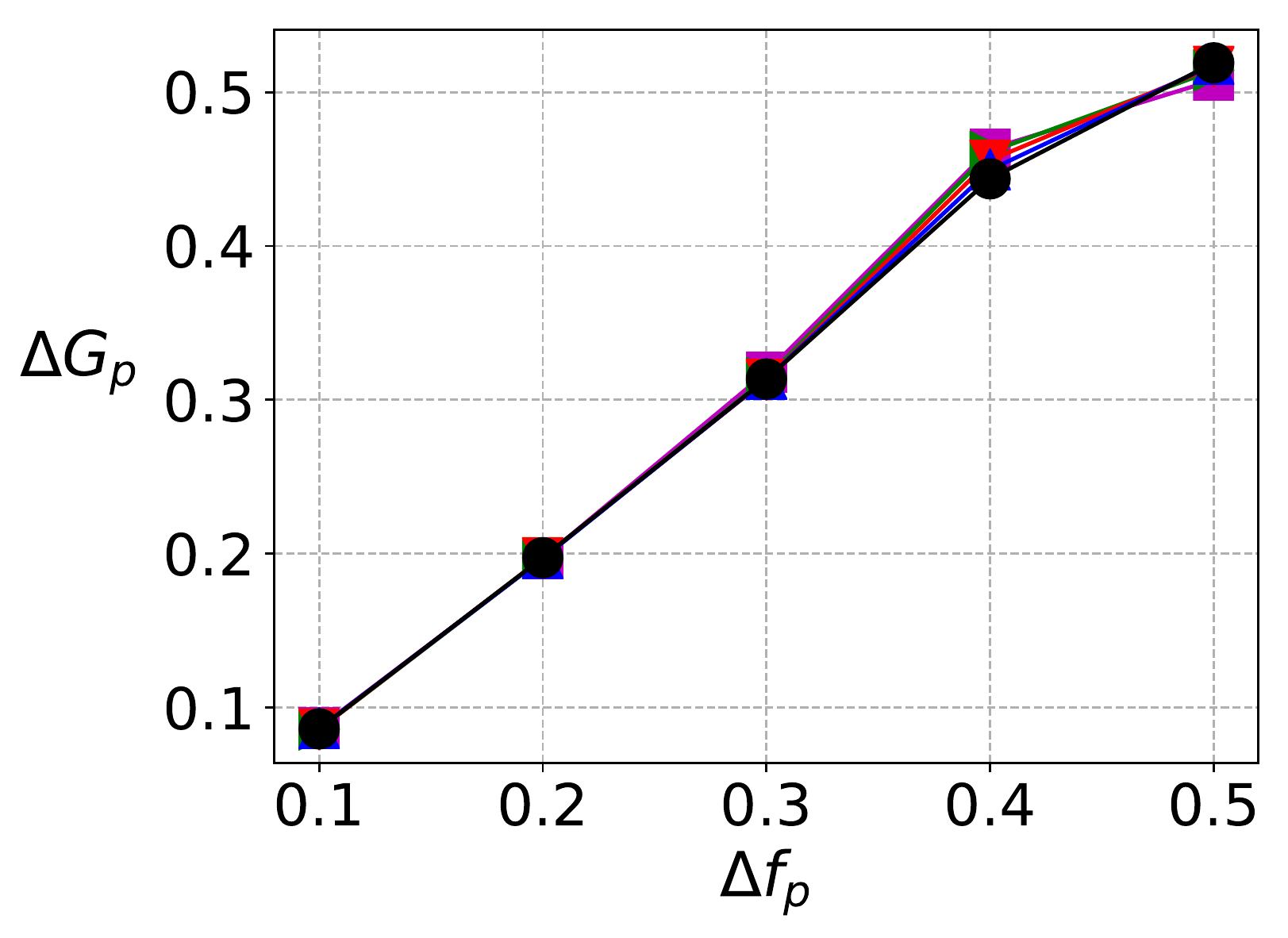}
	\includegraphics[height=4.45cm]{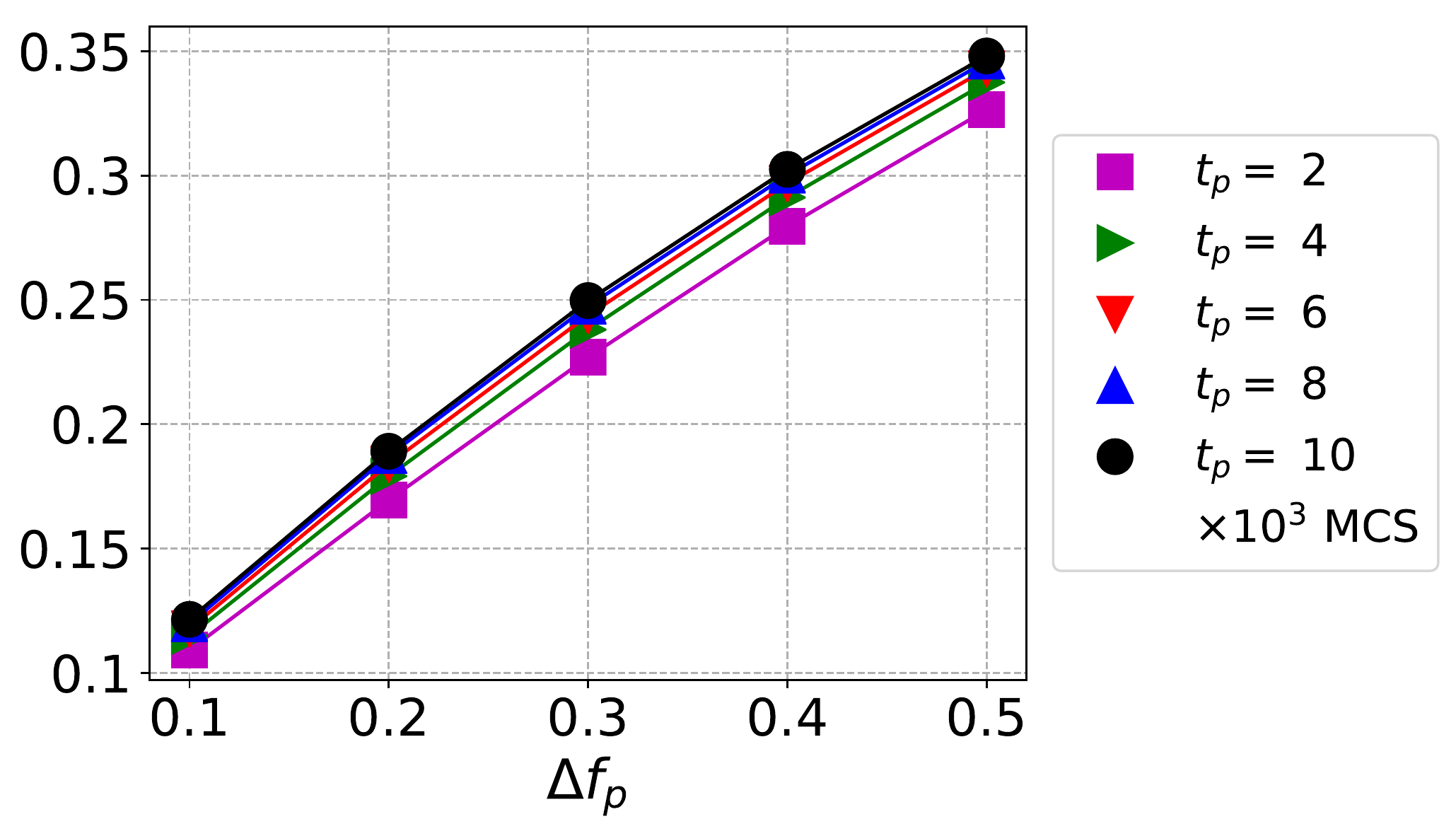}	
	\caption{Difference in the Gini index between states before and immediately after the perturbation, for different values of $\Delta f_{p}$ and $t_p$, with {\it fair} (left) and {\it loser} (right) rule. Changes in the Gini indexes are almost independent of the duration of the perturbation, and bigger for the {it fair} rule. Notice that the vertical axis are different for the two rules.}
	\label{fig:deltaGp}
\end{figure}

A first measure of the short term impact is the difference in the Gini index between states before and immediately after the perturbation, that is, $\Delta G_p = G(t_p) - G(0)$. It is shown on Fig.~\ref{fig:deltaGp} that, for both rules, the greater the value of $\Delta f_p$, the greater that difference, as expected. Due to the slower relaxation of {\it loser} rule, as shown in Fig.~\ref{fig:mse}, that difference increases a little when increasing the perturbation time. Reciprocally, in the {\it fair} rule, that difference is almost constant in terms of perturbation time.

After the perturbation, the time series of Gini index has a stretched (or compressed, as we will see below) exponential relaxation: $$G(t \geq t_p) = G_{\infty} + A \exp \left [-\left(\frac{t - t_p}{\tau} \right)^\alpha \right],$$
where $G_{\infty}$ is the equilibrium Gini index after perturbation, $\tau$ is a characteristic relaxation time and $\alpha$ is a positive exponent. This exponent is called {\it stretching} exponent if it is lower than $1$, and {\it compressing} exponent if it is greater than $1$. Such functional form indicates that, after the perturbation period, a new equilibrium state is attained. This fact, however, does not guarantee that the Gini index of the new equilibrium state is equal to the initial, pre perturbation Gini index; in other words, the perturbation may be irreversible. We define $\Delta G_{irr} = G_{\infty} - G_0$ the difference between the original and the final gini indexes, that measures, if different from zero, the irreversibility of the Gini index after the perturbation process. 

\begin{figure}[!htb]
	\centering
	\includegraphics[height=4.45cm]{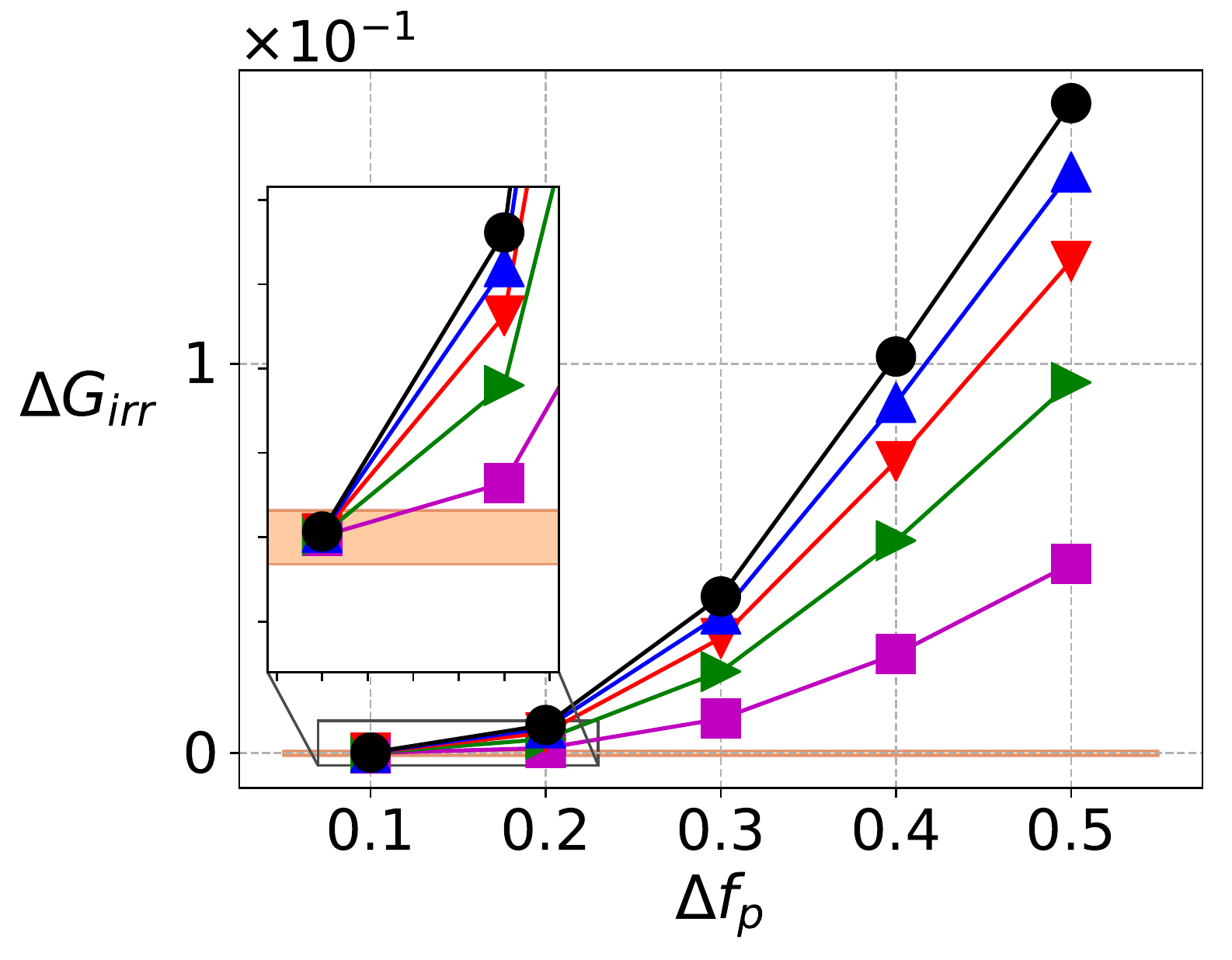}
	\includegraphics[height=4.45cm]{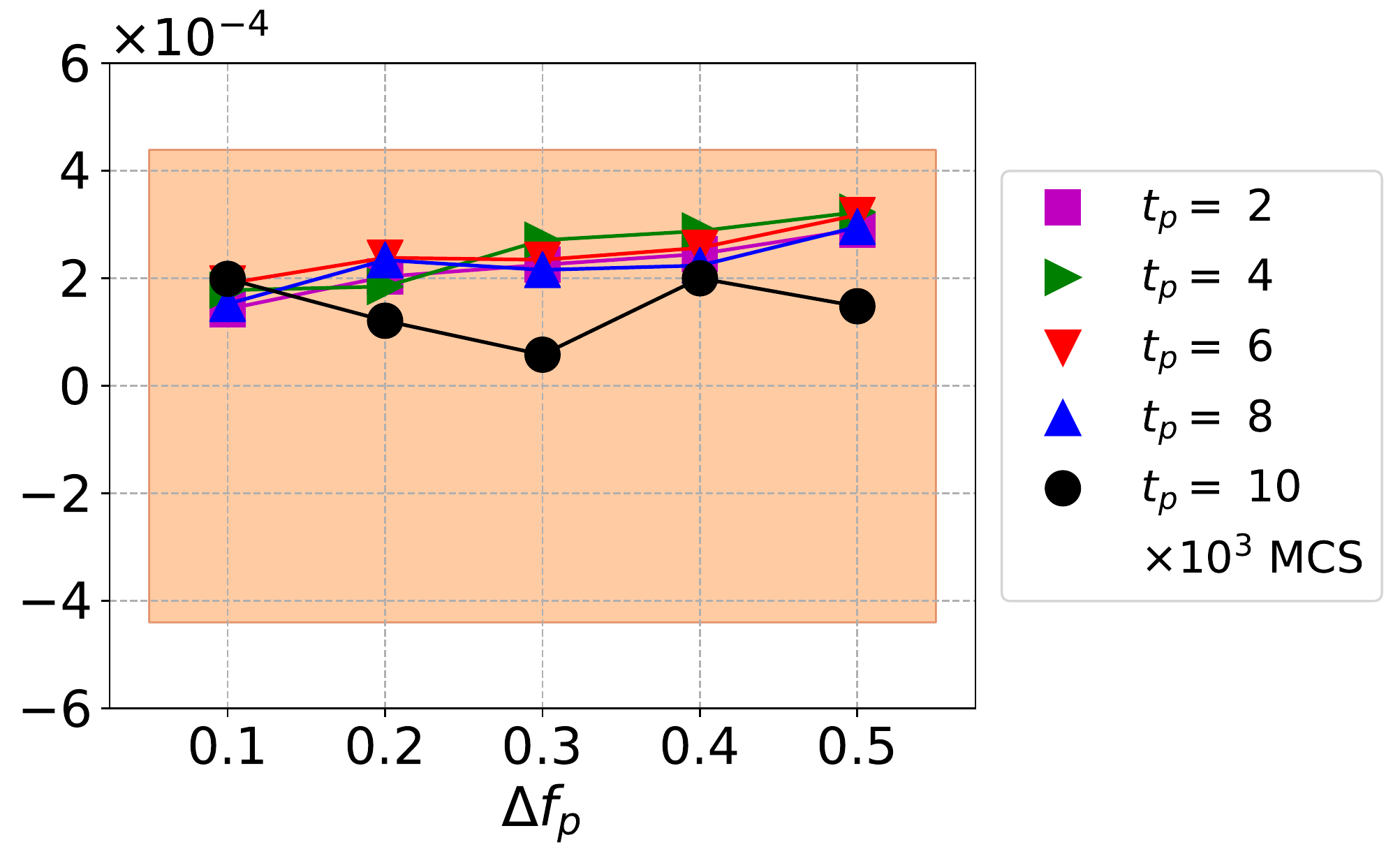}	
	\caption{Difference in the Gini index between equilibrium states before and after the perturbation, for different values of $\Delta f_{p}$ and $t_p$, with {\it fair} (left) and {\it loser} (right) rule. Note the differences in the two scales. The colored area represents the error bar.}
	\label{fig:deltaG}
\end{figure}

We show on Fig.~\ref{fig:deltaG} the cases in which $\Delta G_{irr} = G_{\infty} - G_0$ is within the margin of error (explained in the previous section). For the {\it loser} rule, it is verified that the final state is statistically equivalent to the initial one, in all cases; for the {\it fair} rule the same occurs for $\Delta f_p = 0.1$. In the other cases, the opposite is perceived: the difference is outside the margin of error and, so, the system is irreversible. The interruption of social protection in this particular rule, therefore, generates greater inequality in any future scenario, which is all the greater the longer the time and the intensity of the perturbation. Technically, the result is easily understood: after the pertubation the wealth of a significant fraction of the agents is nullify and, in a similar way as it was shown in Fig.~\ref{fig:zero_time},  they can not get out of this situation, independent of any further change of $f$.

\begin{figure}[!htb]
	\centering
	\includegraphics[height=4.45cm]{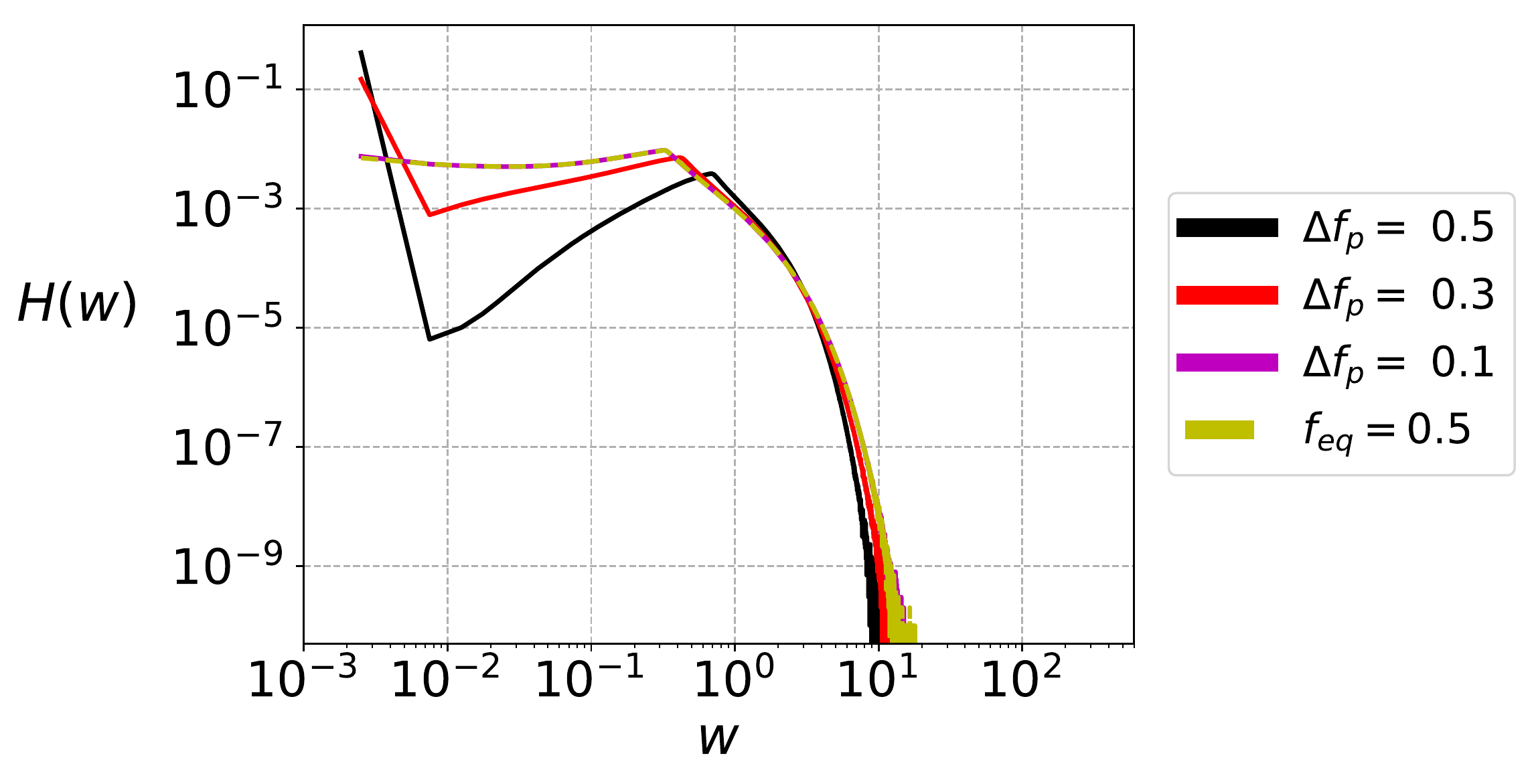}
	\caption{Post-perturbation wealth distributions for the {\it fair} rule. $H(w)$ are the equilibrium distribution after different perturbations $\Delta f_p$ applied during $t_p = 6 \times 10^3 MCS$. All of them averaged over $10^4 MCS$. Note the reference distribution of the initial pre-perturbation equilibrium wealth distribution with $f=0.5$.}
	\label{fig:distribution_pert}
\end{figure}

In order to deeper inspect the irreversible situations, we show in Fig.~\ref{fig:distribution_pert} the distribution functions $H(w)$ for the {\it fair} rule, after perturbations applied for $t_p = 6\times 10^3 MCS$.
Clearly, the irreversibility of the Gini index is associated with the decreasing density of the middle-wealth class and the correspondent increase in the lower-wealth region. This happens if $\Delta f_p > 0.1$ and the effect is more dramatic the greater the perturbation.
The shift of agents from the middle class to the lower class is something similar to what is obtained with a different process~\cite{laguna2005economic}, where agents do not interact if their wealth are apart in more than a certain gap. Nonetheless, here, the result comes from the irreversibility caused by the temporary suspension of a protection policy.
If, however, $\Delta f_p \lesssim 0.1$,  the wealth distributions return to the pre-perturbation distributions. Similar results are obtained for other values of $t_p$ and are not represented here.

For the {\it loser} rule, the post-perturbation equilibrium wealth distributions are equivalent to the pre-perturbation one in all cases, as expected because of the reversibility of the Gini index, so we do not show them.

\subsection{Transient behavior}
We have shown that the system arrives to an equilibrium state in all situations studied. However, the behavior of the transient between the perturbation and the new equilibrium depends, for each of the studied rules, on the intensity and duration of the perturbation. That takes us to the problem of finding or defining an adequate relaxation time. Nonetheless, for the systems studied here, a practical interpretation of the parameters $\tau$ and $\alpha$ makes sense only from a mathematical point of view, as showed in the Appendix~\ref{appendix}. In a real system, a slow relaxation may maintain a lower inequality than a faster one, if the first perturbation has a lower $\Delta G_p$. So, it is important to define a characteristic time that takes this into consideration. With this objective in mind, we define a factor $1-X$ that is the relation between $G(t) - G_{\infty}$ and $G_{\infty}$, that is: 
$$G(t_X) - G_{\infty} = (1 - X)G_{\infty} \Rightarrow t_X = \tau \left[\log \left( \frac{A}{(1 - X)G_{\infty}}\right) \right]^{1/\alpha}$$

\begin{figure}[!htb]
	\centering	
	\includegraphics[height=4.45cm]{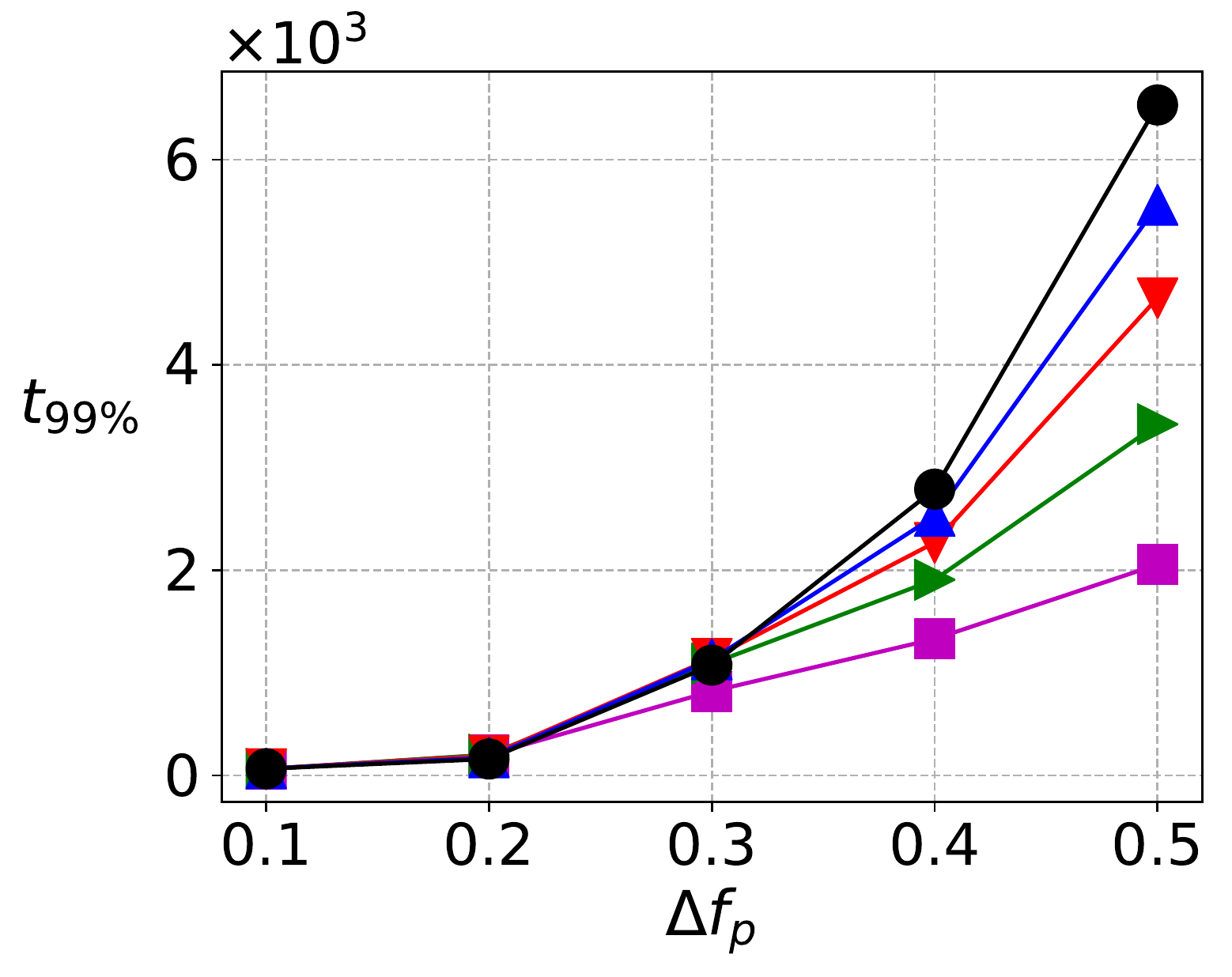}
	\includegraphics[height=4.45cm]{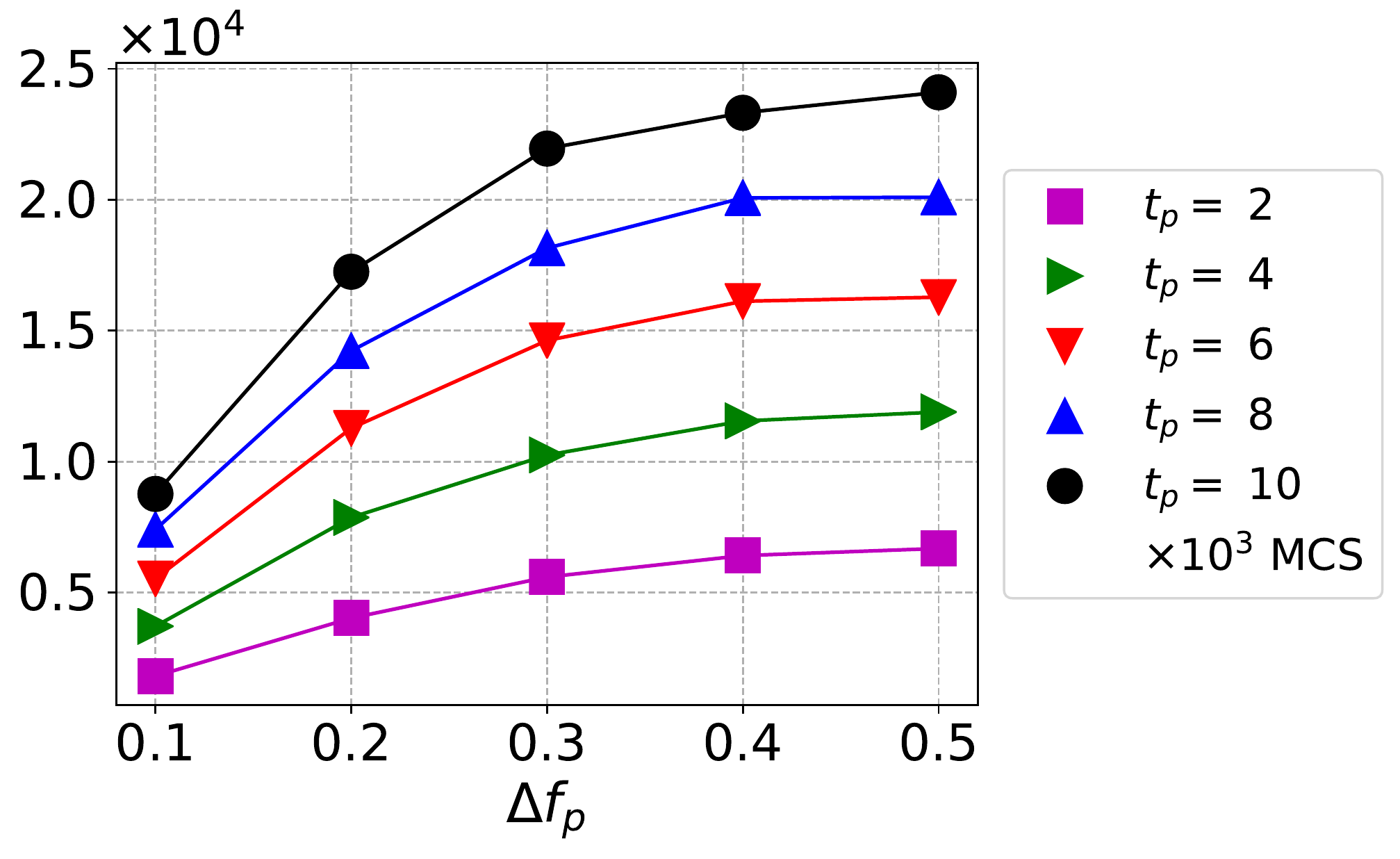}	
	\caption{Characteristic relaxation time for different values of $\Delta f_{p}$ and $t_p$, with {\it fair} (left) and {\it loser} (right) rule. Notice the multiplicative factor for the vertical axis are different for the two rules}
	\label{fig:t99}
\end{figure}

\begin{figure}[!htb]
	\centering	
	\includegraphics[height=4.45cm]{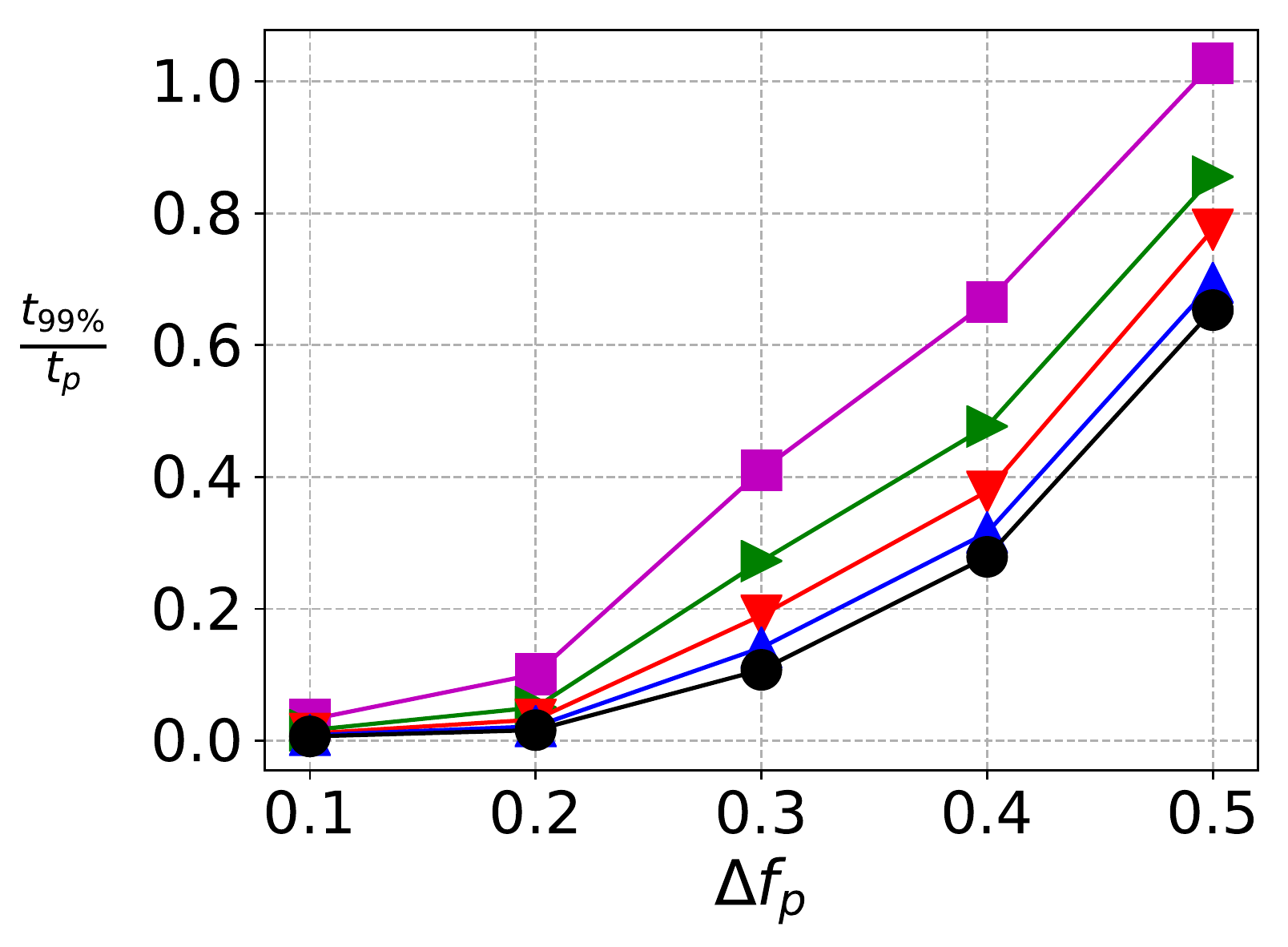}
	\includegraphics[height=4.45cm]{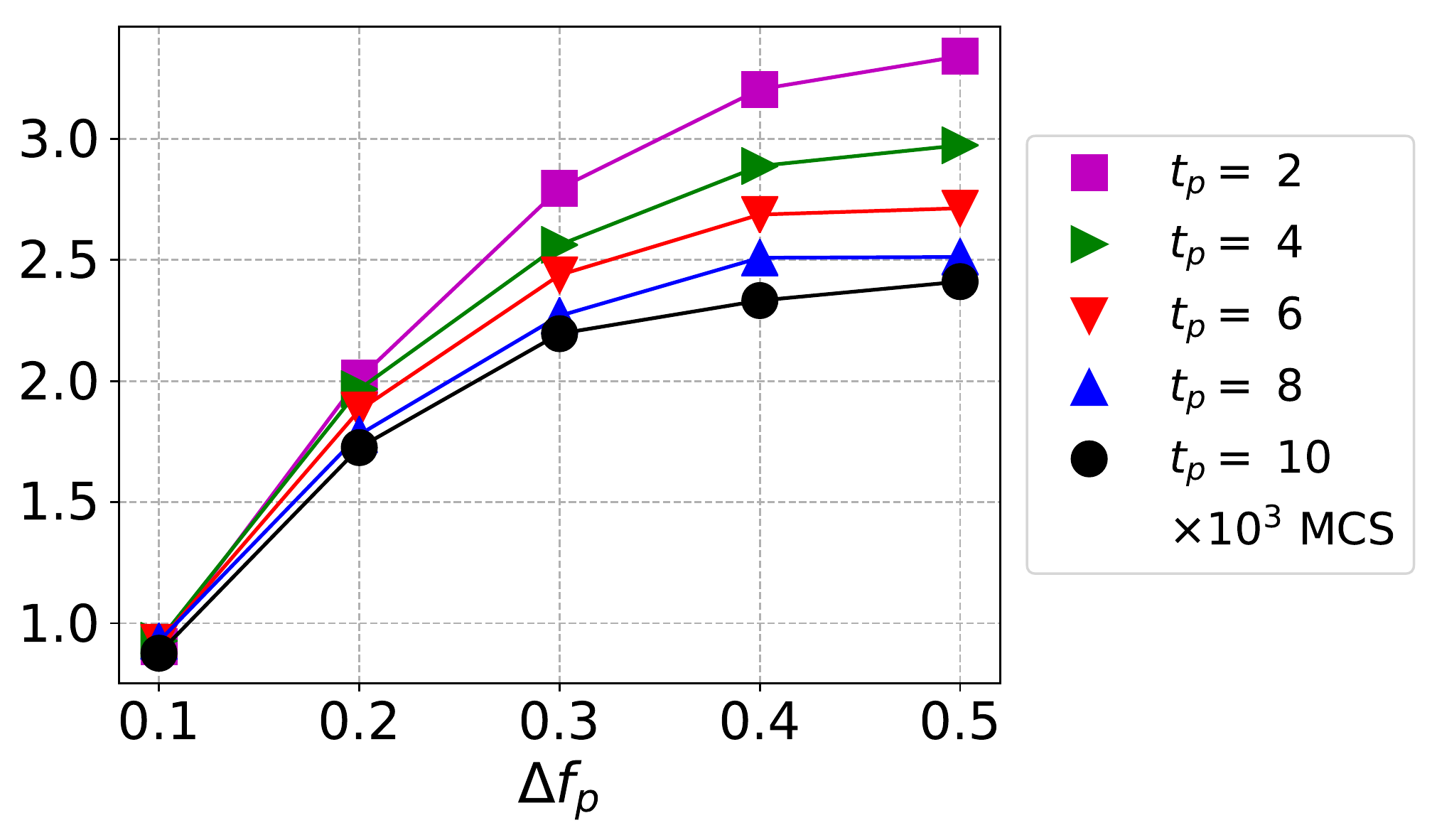}	
	\caption{Characteristic relaxation time and perturbation time ratio for different values of $\Delta f_{p}$ and $t_p$, with {\it fair} (left) and {\it loser} (right) rule.}
	\label{fig:t99_div}
\end{figure}

This means that definying a value of $X$, for example $0.9$ we obtain a characteristic time $t_X$ for which the Gini coefficient is $10\%$ higher than the equilibrium Gini value $G_{\infty}$. In Fig.\ref{fig:t99} and in order to study the system very near equilibrium,  we use $X = 99\%$. It is possible to observe that the grater the pertubation time and the value  $\Delta f_p$, the grater the characteristic time, {\em i.e.} the time to arrive to equilibrium. In relation to the perturbation time, as it is shown in Fig.\ref{fig:t99_div}, the lower the pertubation time and the greater the value of $\Delta f_p$, the greater the relation $\frac{t_{99\%}}{t_P}$. Comparatively, the {\it fair} rule relaxation is faster in almost all cases.

\section{Conclusions}

We have presented here a detailed study of the approach to equilibrium in a system of interacting agents, where two of them are sequentially and randomly chosen to exchange an amount of wealth according to two rules, called {\it fair} and {\it loser} rule. The first rule states that the exchanged quantity is the minimum of the two risked wealths; the second one states that this quantity is equal to the wealth risked by the loser agent. One important factor in the exchange is the factor $f$ in Eq.~\ref{eq:sca} that allows to improve the chances of winning for the poorer agent in each exchange, being this factor an enhancement factor of the difference of wealth between the two agents. The factor $f$ represents then a regulation, or social protection mechanism. This kind of protection may change according to the political situation of a country. Thus, we have also investigated the effect of changes, or simply suppression of this social protection, and how much time the system needs to return to the original situation when social protection is restored, even if in some situations it does not return at all.

We have demonstrated that, for both rules, systems with a constant $f$ always arrive to an stable or equilibrium state. More important, we have verified that a strong and constant social protection policy induce an active redistribution of wealth and this result in a positive influence in the economy, increasing liquidity, trade, exchanges and economic mobility, and reducing inequality. The {\it loser} rule, even if looking "unfair", helps to reduce inequality in all situations, with the eventual exceptions of a  social protection factor bigger than $0.3$. 

Remarkable, our results clearly show that the interruption of social protection entails a high cost associated with the hysteresis of the distribution of wealth. For both rules, the longer the perturbation period and the greater the perturbation intensity, the greater the inequality induced in the short-term. In the middle and long term, the consequences are reciprocal for the two rules. Regarding the perturbation times, the {\it fair} rule presents a fast relaxation, but the inequality is irreversible in most cases, in such a way that the greater the perturbation time and intensity, the greater the irreversibility. In the case of the {\it loser} rule, all cases are reversible, but accompanied by a slower relaxation.

These results are valid for different initial conditions providing that no finite fraction of agents have zero wealth. Also, the values of the risk aversion are constant all along the simulations. Different rules and a change in time of the risk aversion, as exemplified in, are subject of future work.

\bibliography{ref}{}
\bibliographystyle{ieeetr}

\appendix
\section{Technical questions about Perturbation}\label{appendix}
\begin{figure}[!htb]
	\centering
	\includegraphics[height=4.45cm]{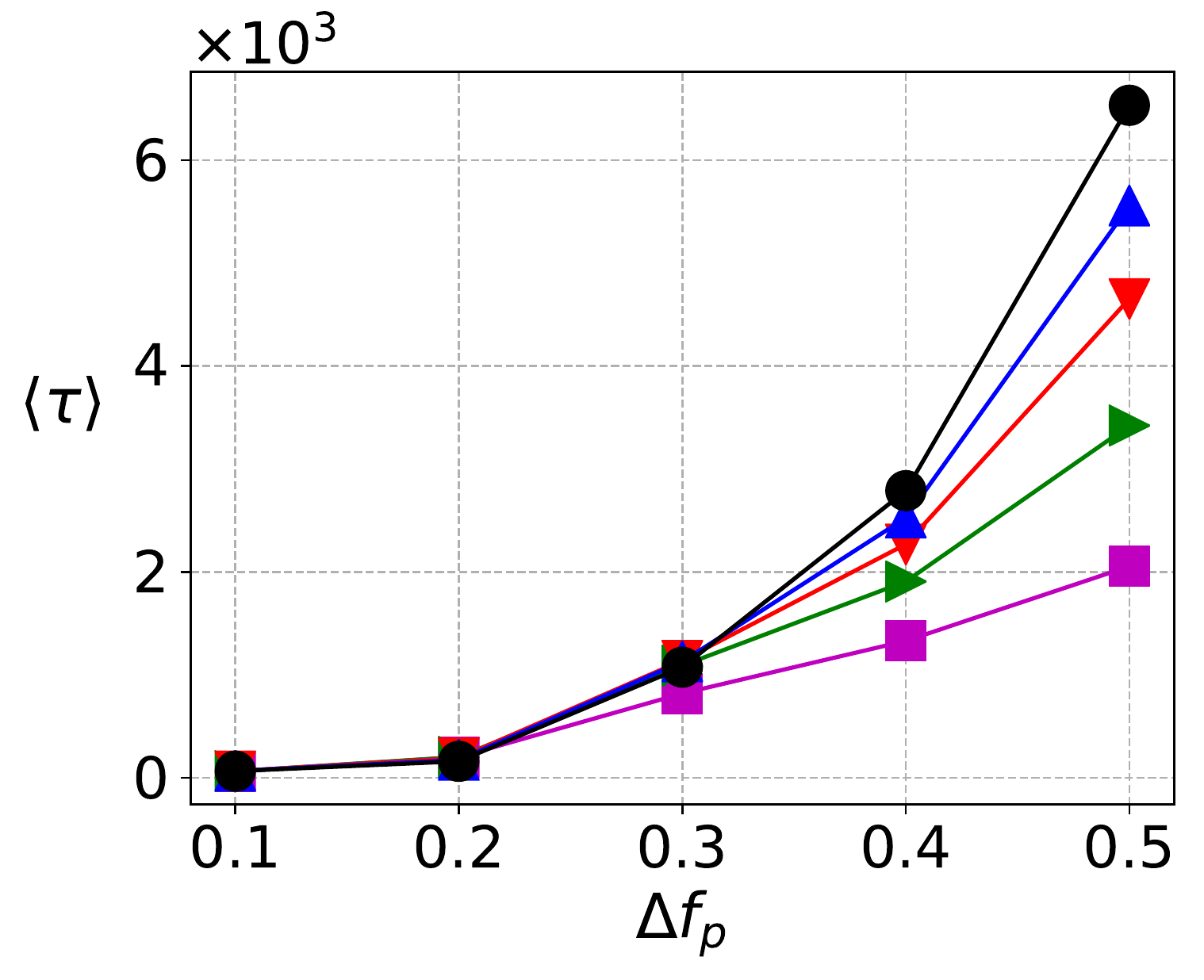}
	\includegraphics[height=4.45cm]{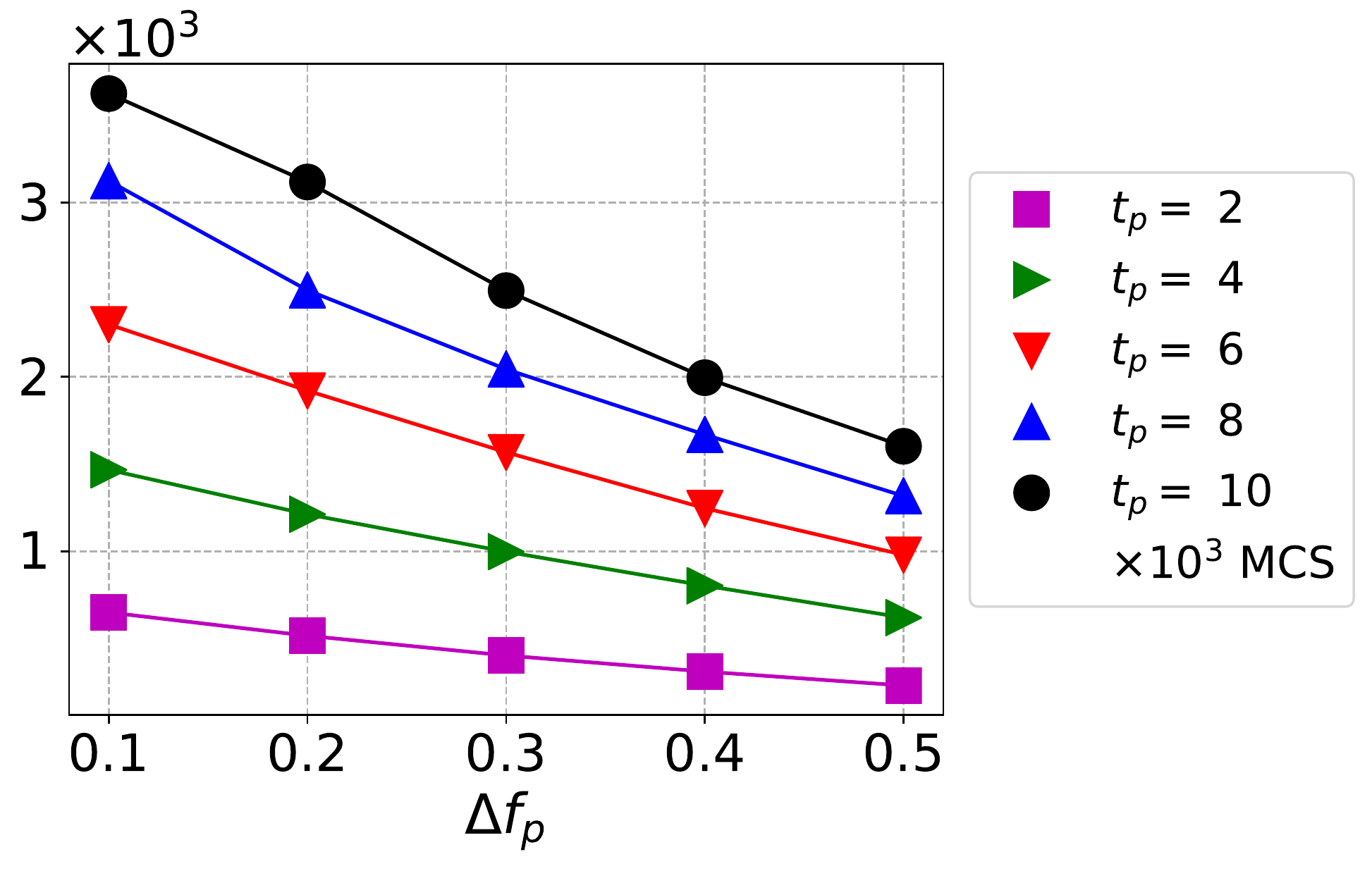}
	\caption{Average relaxation time for different values of $f_{p}$ and $t_p$, with {\it fair} (left) and {\it loser} (right) rule.}
	\label{fig:tau_ave}
\end{figure}

In fact, the systems equilibrate in all situations studied. However, such says little about the behavior of the transient between the perturbation and the new equilibrium state. A first general measure about that behavior is the average relaxation time, given by~\cite{ gradshteyn2014table, bouchaud2008anomalous}: 
$$\langle \tau \rangle = \int_0^{\infty}\exp\left[-\left(\frac{t}{\tau}\right)^{\alpha}\right]dt = \frac{\tau}{\alpha}\Gamma\left(\frac{1}{\alpha}\right),$$ 
where $\Gamma(x)$ is the Gamma function. We show on Fig.~\ref{fig:tau_ave} that, with the exception of $\Delta f_p = 0.5$, the {\it loser} rule relaxation is slower than the {\it fair} rule one. Generally, for both rules, the longer the pertubation time, the slower the relaxation. The influence of the intensity of perturbation, however, is different for the two rules. For the {\it fair} rule, there is a non trivial behavior: the relaxation is very fast with $f_p \geq 0.2$, and, in the other cases, the greater the value of $\Delta f_p$, the slower the relaxation. For the {\it loser} rule, on the other hand, the slower the value of $\Delta f_p$, the slower the relaxation, in all cases.

\begin{figure}[!htb]
	\centering
	\includegraphics[height=4.45cm]{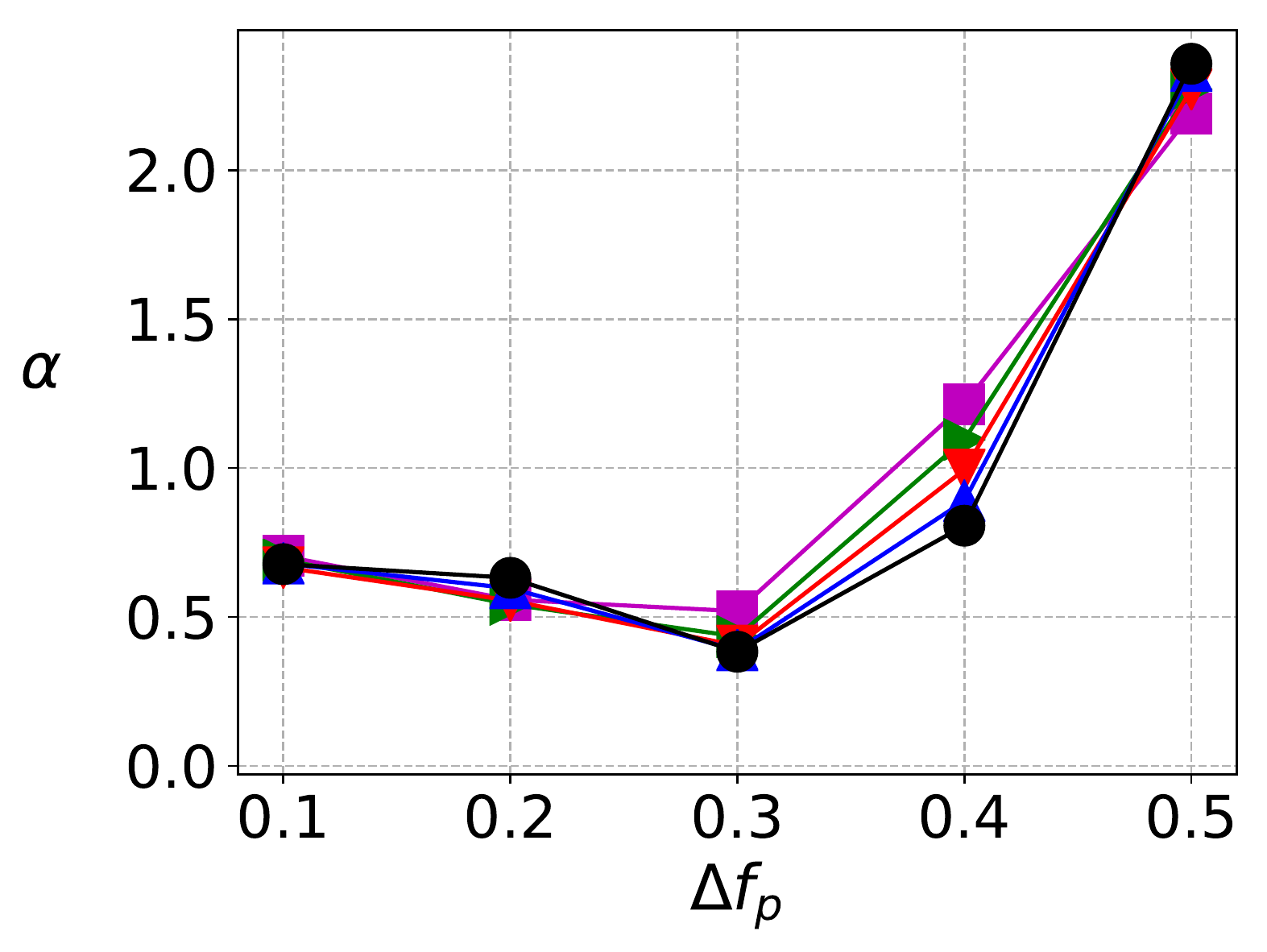}
	\includegraphics[height=4.45cm]{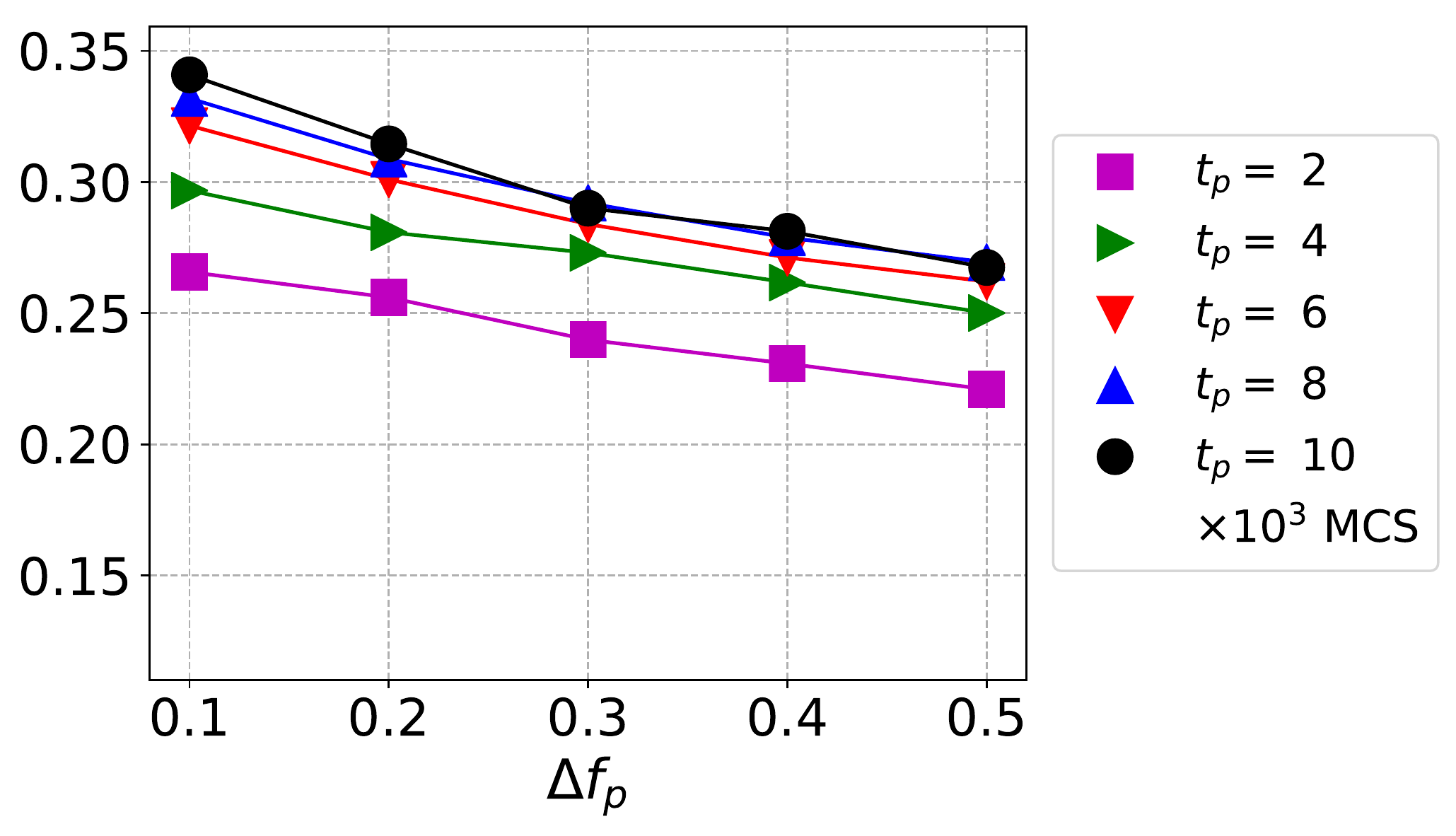}
	\caption{Stretching / compressing exponent for different values of $f_{pert}$, with {\it fair} (left) and {\it loser} (right) rule.}
	\label{fig:alpha}
\end{figure}

In comparison with the exponential $\exp(-t/\langle \tau \rangle)$, the short and long term regimes in the transient may be understood with two reciprocal behavior~\cite{bouchaud2008anomalous}: one such that relaxation is faster and one such that the relaxation is slower. Here we focus on the short-term relaxation, where the long-term one is implicit, since it has the reciprocal behavior. In the stretching case ($\alpha < 1$), the lower the value of $\alpha$, the faster the relaxation on short times. This can be observed in Fig. \ref{fig:alpha}, where, for the {\it loser} rule, the greater the perturbation time and the intensity of perturbation, the slower the relaxation on short times. Also, this can be a good approximation for the {\it fair} rule, in most cases with $\Delta f_p \leq 0.4$. Analogously, in the compressing case ($\alpha > 1$), the higher the value of $\alpha$, the slower the relaxation on short times. This is the case for the {\it fair} rule with $\Delta f_p = 0.5$ and in some cases with $\Delta f_p = 0.4$, where the first one have a faster relaxation on short times than the second one.

\end{document}